\documentclass[preprint,12pt]{elsarticle}

\usepackage{amssymb}
\usepackage{amsmath}
\usepackage{hyperref}

\usepackage[ruled,vlined]{algorithm2e}

\usepackage{algorithmic}

\usepackage{float}
\RestyleAlgo{ruled}
\usepackage{float}
\usepackage{xcolor}  
\colorlet{red}{black}
\colorlet{blue}{black}
\usepackage{amsmath}
\usepackage{multirow}
\usepackage{makecell}
\usepackage{pifont} 
\usepackage{booktabs} 
\usepackage{arydshln} 
\usepackage{bigstrut}    
\usepackage{subfig}
\usepackage{microtype}

\journal{Applied Energy}

\begin{document}

\begin{frontmatter}

\title{Universal Transient Stability Analysis: A Pre-trained Generative Transformer-Enabled Power System Dynamics Prediction Framework}

\author[label1]{Chao Shen}
\author[label1]{Ke Zuo}
\author[label2]{Mingyang Sun \texorpdfstring{\corref{cor1}}{}}
\cortext[cor1]{Corresponding Author}

\affiliation[label1]{organization={College of Control Science and Engineering, Zhejiang University},
            postcode={310027},
            city={Hangzhou},
            country={China}}

\affiliation[label2]{organization={School of Advanced Manufacturing and Robotics,Peking University},
            postcode={100871},
            city={Beijing},
            country={China}}

\begin{abstract}
Existing dynamics prediction frameworks for transient stability analysis (TSA) fail to achieve multi-scenario "universality"—the inherent ability of a single, pre-trained architecture to generalize across diverse operating conditions, unseen faults, and heterogeneous systems. To address this, this paper proposes Uni-TSA, a pre-trained generative Transformer-enabled universal framework that models multivariate transient dynamics prediction as a univariate generative task with three key innovations: First, a novel data processing pipeline featuring channel independence decomposition to resolve dimensional heterogeneity, sample-wise normalization to eliminate separate stable/unstable pipelines, and temporal patching for efficient long-sequence modeling; Second, a parameter-efficient freeze-and-finetune strategy that augments the pre-trained generative Transformer backbone with dedicated input embedding and output projection layers while freezing core transformer blocks to preserve generic feature extraction capabilities; Third, a two-stage fine-tuning scheme that combines teacher forcing, which feeds the model ground-truth data during initial training, with scheduled sampling, which gradually shifts to leveraging model-generated predictions, to mitigate cumulative errors in long-horizon iterative prediction. Comprehensive testing demonstrates the framework's universality, as Uni-TSA trained solely on the New England 39-bus system achieves zero-shot generalization to mixed stability conditions and unseen faults, and matches expert performance on the Iceland 189-bus system with only 5\% fine-tuning data. Additional cross-system experiments on the IEEE 68-bus and IEEE 118-bus systems, together with stability metrics and PEBS comparison, further confirm Uni-TSA's strong zero-shot transferability and data-efficient adaptation. This multi-scenario versatility validates a universal framework that eliminates scenario-specific retraining and achieves scalability via large-scale parameters and cross-scenario training data.
\end{abstract}

{\color{black}

\begin{highlights}
  \item A pre-trained generative Transformer-enabled TSA framework enables universal trajectory prediction.
  \item Channel independence, normalization, and patching enable system-agnostic modeling.
  \item Freeze-and-finetune GPT preserves pre-trained sequence feature extraction for TSA.
  \item TeaF-SchS fine-tuning mitigates error accumulation in autoregressive rollout.
  \item Zero-shot and few-shot tests show transfer across faults, OCs, and systems.
  \end{highlights}
}

\begin{keyword}
Deep learning, data-driven methods, transient stability analysis, dynamic trajectory prediction, pre-trained generative Transformer, iterative prediction.
\end{keyword}

\end{frontmatter}




\section{Introduction}

The increasing penetration of renewable energy sources reduces system inertia and elevates operational uncertainty, posing significant challenges to grid stability \cite{shen2025physics, segurado2011increasing}. Consequently, modern power systems are increasingly susceptible to contingencies, demanding more robust analytical frameworks. {\color{red}Power system stability is the ability of an electric power system, for a given initial operating condition, to regain a state of operating equilibrium after being subjected to a physical disturbance, with most system variables bounded so that practically the entire system remains intact \cite{transient_stability_definition}.} A key aspect of this broad stability notion is transient stability analysis (TSA), which {\color{red}assesses whether the system can maintain synchronism and reach an acceptable post-disturbance equilibrium following a large disturbance (e.g., short-circuit faults, line/generator tripping), as manifested in the post-disturbance transient dynamics} \cite{shi2020convolutional, ye2024use, transient_stability_definition}. {\color{red}Accordingly, the core objective of this work is to support TSA by rapidly predicting post-fault trajectories of key state variables (e.g., rotor angle and speed) from limited PMU observations, thereby enabling timely stability assessment and emergency decision-making.} 

TSA is critical for preventive control and emergency response, enabling grid operators to enhance resilience against cascading failures and ensure reliable operation under evolving operating conditions (OCs) \cite{zhao2019exergy}. Conventionally, the dynamic behavior of power systems is {\color{red}typically described by} nonlinear differential-algebraic equations (DAEs), which are numerically solved via time-domain simulation (TDS) methods \cite{zhan2023hybrid}. While TDS provides high-fidelity temporal resolution, its computational intensity and scalability challenges limit its online practicality \cite{zhao2022structure}. 
{\color{red}
Alternatively, {direct methods} evaluate transient stability without performing long-horizon time-domain simulation. Following the taxonomy summarized in \cite{qiu2022adaptive}, mainstream direct methods can be categorized into three groups. (i) {EAC-based equivalencing methods} (e.g., the Extended Equal Area Criterion (EEAC) \cite{xue1989extended}) extend the equal-area criterion from SMIB to multi-machine systems by coherent-grouping and two-machine equivalencing \cite{kundur2007power}. In practice, they often become hybrid integral-direct methods due to the need for trajectory-dependent grouping or auxiliary integration. (ii) {Global energy-function (topological) methods} \cite{chang1995direct} construct a global energy function $V(x)$ and characterize the stability region via the closest controlling unstable equilibrium point (UEP) \cite{berggren2002nature}. A representative practical variant is the Potential Energy Boundary Surface (PEBS) method, which approximates the stability boundary by locating the first local maximum of the potential energy along the fault-on trajectory, thereby avoiding explicit UEP computation \cite{chiang1988pebs}. While these methods provide clear geometric interpretation and stability margins, they rely on strong model assumptions (existence of a global energy function) and typically still require obtaining the fault-clearing state. (iii) {Local Lyapunov-function (LLF) methods \cite{qiu2022adaptive, zong2023transient, willems2007application}} construct a Lyapunov function locally around the post-fault stable equilibrium and certify stability via an invariant sublevel set, which can, in principle, avoid numerical integration and UEP computation. However, traditional LLF conditions and set/derivative bounds may introduce conservatism.}
{\color{red}Despite their theoretical elegance, direct methods face challenges in model generality, numerical tractability, and conservatism in large-scale online settings \cite{qiu2022adaptive, azman2020unified}.}

The widespread deployment of phasor measurement units (PMUs) has spurred the development of data-driven TSA. These methods learn the mapping between system measurements and stability-related features, enabling rapid online applications such as real-time stability classification \cite{zhou1994application,gomez2010support,Sun2007An,chen2022interpretable,ren2021interpretable,azman2020unified,zhu2023integrated} and stability margin estimation \cite{su2023online,zhu2019hierarchical,zhu2021networked}. Early research in this domain was dominated by classical machine learning. For instance, artificial neural networks were applied to perform rapid transient stability classification \cite{zhou1994application}, while support vector machines demonstrated robust post-fault stability prediction using synchrophasor data \cite{zou2020least, gomez2010support}. Concurrently, decision trees were used to extract critical stability indicators through offline feature ranking, facilitating periodic updates to security rules \cite{li2022deep, Sun2007An}. More recent research has shifted toward deep learning (DL) frameworks to improve both prediction accuracy and model interpretability. To enhance explainability, a time-adaptive attention-based gated recurrent unit (GRU) was developed to visualize feature importance \cite{chen2022interpretable}, while other work integrated decision tree path length constraints into the DL loss function \cite{ren2021interpretable}. For performance improvement, hybrid architectures combining convolutional neural networks (CNNs) and long short-term memory (LSTMs) were proposed for stability classification and oscillation prediction \cite{azman2020unified, shi2020convolutional}. Beyond binary classification, significant effort has focused on directly predicting stability margins, often derived from the critical clearing time (CCT). Notable methods include an ensemble DL framework with dynamic error correction for CCT estimation \cite{su2023online} and a two-level CNN regression model that predicts margins from 2-D images of system trajectories \cite{zhu2019hierarchical}.

Although existing data-driven models provide rapid and accurate stability classification, this binary output offers limited actionable insight for system operators \cite{zhao2022structure, ye2024use}. Consequently, TSA research has shifted toward predicting detailed dynamic trajectories, which are critical for informing corrective actions such as load shedding or out-of-step (OOS) SG tripping \cite{ye2024use}. One prominent approach involves physics-informed neural networks (PINNs), which incorporate physical constraints as regularization terms to predict system dynamics \cite{misyris2020physics}. However, the applicability of early PINN models was often limited to stable OCs. Subsequent work on physics-following neural networks (PFNNs) has sought to improve generalization across both stable and unstable cases by introducing time-varying algebraic parameters and supervised initialization into the PINN framework \cite{shen2025physics}. Other notable strategies include using LSTMs with separate processing pipeline for stable and unstable samples \cite{ye2024use} and leveraging Fourier neural operators (FNOs) to predict transients in the frequency domain, thereby simplifying the learning task \cite{cui2023frequency}.

{\color{black}
While the methods in \cite{shen2025physics, ye2024use, cui2023frequency} improve prediction accuracy in mixed stable/unstable OCs, they exhibit poor generalization to unseen faults. To address this, the stochastic variational deep kernel regressor (SVDKR)-based framework in \cite{tan2025bayesian} integrates an OOS detector for classifying SGs as stable or unstable. Dedicated SVDKR predictors are trained per stability class, followed by per-SG error correctors to refine unit-level predictions. Although this improves robustness, the parameter complexity and computational overhead hinder its scalability to large systems  with numerous SGs. The deep neural representation (DNR) framework was proposed to mitigate these issues \cite{zhao2022structure}, using a unified graph attention and GRU architecture to predict dynamics in a single model. This design enhances performance under unseen faults but suffers from significant cumulative error amplification over long prediction horizons. More importantly, both the SVDKR \cite{tan2025bayesian} and DNR \cite{zhao2022structure} frameworks are designed for a single system. {\color{black}They lack the ability to generalize across heterogeneous power systems with different topologies and SG configurations, necessitating laborious retraining for each new system. This fundamental limitation—the absence of cross-system generalization capability—represents a critical knowledge gap in current TSA research, preventing the development of universal frameworks that could enable efficient knowledge transfer across diverse power networks.}
}

{\color{black}
As demonstrated in \autoref{Universality Comparison of the Existing and Proposed Methods}, existing frameworks suffer from pronounced performance degradation across diverse OCs, faults and system configurations. {\color{black}This systematic failure across multiple dimensions reveals a fundamental knowledge gap: \textit{no existing TSA framework has achieved true universality}.} This shortcoming motivates the pursuit of \textit{universality} in TSA models, {\color{black}defined as a novel paradigm representing} the intrinsic capability of a single pre-trained architecture to generalize robustly across mixed stable/unstable OCs, unseen fault events, and heterogeneous power systems. Universality enables a paradigm shift in TSA, analogous to the versatility of large language models (LLMs) in natural language processing, where a unified model adeptly addresses diverse dynamic trajectory prediction tasks \cite{achiam2023gpt}. By eliminating the need for scenario-specific retraining, a universal TSA framework provides foundational capabilities, most prominently exemplified by cross-system generalization. This feature enables efficient knowledge transfer from source-domain systems to large-scale target grids via few-shot adaptation, thereby minimizing computational simulation and labor costs associated with custom model development for diverse power networks \cite{tu2024powerpm}. Furthermore, universality inherently supports scalability, with performance systematically improving as model parameters and multi-scenario training datasets expand, which ensures adaptability to evolving grid complexities \cite{liu2024timer}.}

{\color{red}
To avoid ambiguity, this paper adopts an operational definition of TSA ``universality'' along three explicit generalization axes: (i) \textit{operating-condition universality}, i.e., robustness to mixed stable/unstable post-fault trajectories under diverse pre-fault operating points; (ii) \textit{fault universality}, i.e., generalization to unseen fault locations/clearing conditions beyond those observed in training; and (iii) \textit{cross-system universality}, i.e., reusing the \emph{same} pre-trained architecture and model interface across heterogeneous networks with different SG configurations, with at most lightweight few-shot adaptation instead of full retraining.}

{\color{black}
However, achieving this universality requires overcoming three fundamental limitations: 1) Data processing rigidity, where current methods are constrained to fixed input dimensionalities based on specific state variables, limiting scalability and requiring complete retraining for system changes or monitored SGs modifications \cite{zhao2022structure}, while frequently necessitating separate processing pipelines for stable/unstable dynamics that introduce computational overhead and misclassification risks \cite{tan2025bayesian, ye2024use}; {\color{red}More importantly for cross-system generalization, the number of monitored SGs typically differs across networks, which directly changes the dimension of system-level stacked trajectories. This creates an inherent input/output dimension mismatch for predictors with fixed multivariate interfaces, and is a key reason why many existing models must be redefined or retrained when transferring to a new system.} 2) Architectural constraints, as prevalent recurrent neural network architectures (LSTMs \cite{ye2024use} and GRUs \cite{zhao2022structure}) suffer from vanishing gradient problems that impede long-range dependency modeling and inherently sequential processing that precludes parallel computation, creating critical bottlenecks for model scalability \cite{fan2019assessment, liu2024timer}; 3) Training and error propagation, where the temporal asymmetry between short observation windows and extended prediction horizons in online TSA, combined with existing training schemes that lack explicit error suppression mechanisms, causes prediction errors to accumulate at each time step during iterative prediction, severely degrading long-term reliability \cite{shen2025physics, karimi2022optimization}.}


{
To address these challenges, this study proposes a pre-trained generative Transformer-enabled universal dynamics prediction framework, Uni-TSA, that {\color{red}fundamentally reformulates multivariate transient dynamics prediction as a univariate generative sequence prediction task by leveraging pre-trained generative Transformer parameters whose modality-agnostic sequence feature extraction and temporal dependency modeling capabilities transfer effectively to power-system dynamics (see \ref{Generic Feature Extraction of Transformer Block} for empirical validation).} The main contributions are:

\begin{enumerate}
     \item {\color{black}To the best of the authors' knowledge, this is the {first work} to demonstrate universality in TSA through a pre-trained generative Transformer-enabled approach, achieving generalization across diverse OCs, unseen faults, and heterogeneous systems by systematically addressing fundamental limitations in data processing, model architecture, and fine-tuning schemes. This work establishes a new research direction by bridging generative sequence modeling and power system dynamics prediction. {\color{red}The challenge-to-design mapping is summarized in Table~\ref{tab:challenge-design}.}}
     \item {\color{black}A novel data processing pipeline is designed and operates sequentially: channel-independence decouples multivariate dynamic sequences into univariate streams, resolving dimensional heterogeneity across different systems and enabling system-agnostic processing; sample-wise normalization standardizes sequence samples via internal statistics, eliminating separate pipelines for stable/unstable OCs; temporal patching tokenizes sequences into semantic patches, enabling efficient long-sequence modeling while capturing high-level temporal patterns. This pipeline design transforms the problem from multivariate regression to univariate sequence generation, enabling direct compatibility with pre-trained generative Transformer architectures.}
     \item {\color{black}A Generative Pre-trained Transformer (GPT) architecture is devised that incorporates dedicated input embedding and output projection alongside causal attention mechanisms for temporal dependency modeling. {This represents the first application of pre-trained transformer architectures to power system dynamics prediction, leveraging transfer learning from large-scale sequence modeling tasks.} The freeze-and-finetune strategy preserves pre-trained sequence feature extraction capability by freezing core transformer blocks (T-blocks) while selectively fine-tuning TSA-specific layers, achieving balance between knowledge retention and task adaptation.}
     \item {\color{black}A two-stage fine-tuning scheme is proposed to mitigate error propagation during iterative and generative dynamics prediction. This training strategy is specifically designed for autoregressive sequence generation in TSA, addressing the unique challenge of error accumulation in long-horizon prediction tasks. The initial teacher forcing (TeaF) stage ensures stable convergence by exclusively feeding ground-truth sequences as input. Subsequently, the scheduled sampling (SchS) stage enhances robustness by increasingly replacing ground-truth inputs with the model's own predictions, thereby compelling the model to correct for self-generated errors.}
\end{enumerate}
}
{\color{red}
\begin{table}[H]
\centering
{\color{red}
\caption{Challenge-to-design mapping}
\label{tab:challenge-design}
\small
\setlength{\tabcolsep}{6pt}
\renewcommand{\arraystretch}{1.25}
\begin{tabular}{@{}>{\raggedright\arraybackslash}p{0.28\linewidth} >{\raggedright\arraybackslash}p{0.66\linewidth}@{}}
\toprule
\textbf{Challenge} & \textbf{Design solution} \\
\midrule
Data processing rigidity &
\makecell[tl]{Channel-independence, sample-wise normalization,\\ temporal patching (Sec.~3.2)} \\
\midrule
Architectural constraints &
\makecell[tl]{GPT with causal attention,\\ freeze-and-finetune strategy (Secs.~3.2--3.3)} \\
\midrule
Training and error propagation &
\makecell[tl]{Two-stage TeaF$\rightarrow$SchS fine-tuning\\ (Sec.~3.3)} \\
\bottomrule
\end{tabular}
}
\end{table}}

\begin{table}[H]
  \centering
  \caption{Universality Comparison of Methods}
  \begin{tabular}{ccccc}
\hline
    \multirow{3}{*}{Method} & \multicolumn{3}{c}{Universality Scenarios} & \multirow{3}{*}{\makecell{Single \\ Model}} \\
    \cline{2-4}
    & \makecell{Stable/Unstable \\ OCs} & \makecell{Unseen \\ Faults} & \makecell{Heterogeneous \\ System} & \\
\hline
    PINN \cite{misyris2020physics} & $\times$ & $\times$ & $\times$ & $\checkmark$ \\
    LSTM \cite{ye2024use} & $\checkmark$ & $\times$ & $\times$ & $\times$ \\
    PFNN \cite{shen2025physics} & $\checkmark$ & $\times$ & $\times$ & $\checkmark$ \\
    FNO \cite{cui2023frequency} & $\checkmark$ & $\times$ & $\times$ & $\checkmark$ \\
    SVDKR \cite{tan2025bayesian} & $\checkmark$ & $\checkmark$ & $\times$ & $\times$ \\
    DNR \cite{zhao2022structure} & $\checkmark$ & $\checkmark$ & $\times$ & $\checkmark$ \\
    \textbf{Uni-TSA} & $\checkmark$ & $\checkmark$ & $\checkmark$ & $\checkmark$ \\
\hline
  \end{tabular}
  \label{Universality Comparison of the Existing and Proposed Methods}
\end{table}

\section{Preliminary and Problem Formulation}
\subsection{Power System Model}
{\color{black}
Considering a power system containing $n_b$ buses, $n_g$ SGs, and $n_l$ loads, the DAEs describing the system's transient dynamics can be formulated as follows \cite{shen2025physics, wen2016allocation}:
\begin{align}
\begin{cases}
&\dot{\mathbf{x}}(t)=\mathbf{f}(\mathbf{x}(t), \mathbf{y}(t), \mathbf{u}(t), \mathcal{F}(t), t),\\
&\mathbf{0}=\mathbf{g}(\mathbf{x}(t), \mathbf{y}(t), \mathbf{u}(t), \mathcal{F}(t), t),
\end{cases}
\end{align}
where $\mathbf{x}(t)\in \mathbb{R}^{n_\mathbf{x}}$ is the vector of dynamic state variables (e.g., rotor angles $\delta_i$ and angular speed $\omega_i$ of $i$-th SG), and $\mathbf{y}(t)\in \mathbb{R}^{n_\mathbf{y}}$ is the vector of algebraic variables (e.g., voltage magnitudes $V_i$ of $i$-th bus). The vector $\mathbf{u}(t)\in \mathbb{R}^{n_\mathbf{u}}$ represents control inputs, such as governor setpoints. The fault set $\mathcal{F}(t)$ denotes the system components experiencing faults. The nonlinear functions $\mathbf{f}(\cdot)$ and $\mathbf{g}(\cdot)$ describe the system dynamics and algebraic constraints, respectively.
}

\subsection{Transformer Attention Mechanism}
\begin{figure}[H]
    \centering
    \includegraphics[width=\linewidth]{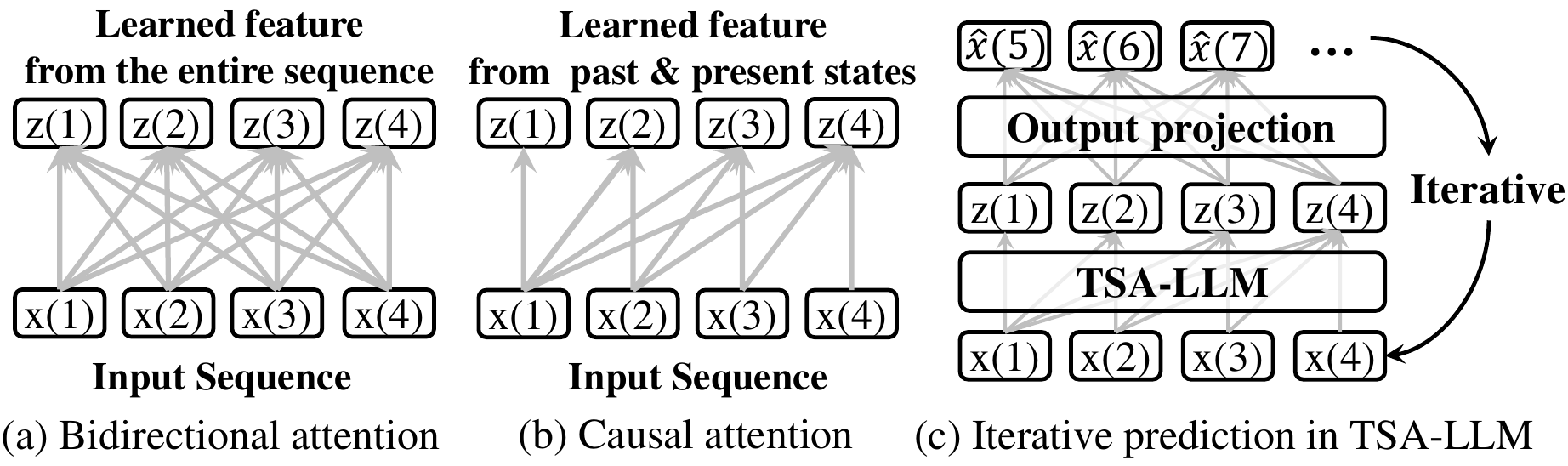}
    \caption{Comparison of attention mechanisms and iterative prediction in Uni-TSA. }
    \label{Comparative analysis of attention mechanisms and prediction paradigms.}
\end{figure}

{\color{black}The transformer architecture has become the dominant backbone of modern large language models \cite{achiam2023gpt, liu2024deepseek, shen2026proopf, shen2026llm} and has demonstrated remarkable success in sequence modeling, owing to its parallelizable computation and ability to capture long-range dependencies \cite{vaswani2017attention, gao2025mmgpt4lf}. At the core of this architecture is the self-attention mechanism. As illustrated in the bidirectional attention mechanism presented in \autoref{Comparative analysis of attention mechanisms and prediction paradigms.}(a), the process begins with an input sequence $\mathbf{X} = [\mathbf{x}(1), \mathbf{x}(2), \ldots, \mathbf{x}(T)] \in \mathbb{R}^{T \times d}$, where $T$ is the sequence length and $d$ is the feature dimension, the bidirectional attention mechanism first transforms the input into three different representations: queries ($\mathbf{Q}$), keys ($\mathbf{K}$), and values ($\mathbf{V}$):
\begin{align}
    \mathbf{Q} = \mathbf{X}\mathbf{W}^Q, \quad \mathbf{K} = \mathbf{X}\mathbf{W}^K, \quad \mathbf{V} = \mathbf{X}\mathbf{W}^V,\label{QKV}
\end{align}
where $\mathbf{W}^Q \in \mathbb{R}^{d \times d_k}$, $\mathbf{W}^K \in \mathbb{R}^{d \times d_k}$, and $\mathbf{W}^V \in \mathbb{R}^{d \times d_v}$ are learnable parameter matrices. The self-attention matrix is then computed as:
\begin{align}
    \text{Attention}(\mathbf{Q}, \mathbf{K}, \mathbf{V}) = \text{softmax}\left({\mathbf{Q}\mathbf{K}^T}/{\sqrt{d_k}}\right)\mathbf{V},
\end{align}
where ${1}/{\sqrt{d_k}}$ is a scaling factor to prevent vanishing gradients during training. In practice, Multi-Head Attention (MHA) is employed to enrich the model's expressive capability:
\begin{align}
    \text{MHA}(\mathbf{X}) = \text{Concat}(\text{head}_1, \text{head}_2, \ldots, \text{head}_h)\mathbf{W}^O,\label{MHA}
\end{align}
where $\text{head}_i = \text{Attention}(\mathbf{X}\mathbf{W}^Q_i, \mathbf{X}\mathbf{W}^K_i, \mathbf{X}\mathbf{W}^V_i)$. $\mathbf{W}^Q_i \in \mathbb{R}^{d \times d_k/h}$, $\mathbf{W}^K_i \in \mathbb{R}^{d \times d_k/h}$, $\mathbf{W}^V_i \in \mathbb{R}^{d \times d_v/h}$ are parameter matrices for the $i$-th attention head, $\mathbf{W}^O \in \mathbb{R}^{d_v \times d}$ is the output projection matrix, and $h$ is the number of attention heads. By operating on the entire sequence, this bidirectional attention mechanism is well-suited for tasks like sequence translation and classification \cite{vaswani2017attention, gao2025mmgpt4lf}.}

\section{Methodology}
\begin{figure}
    \centering
    \includegraphics[width=\linewidth]{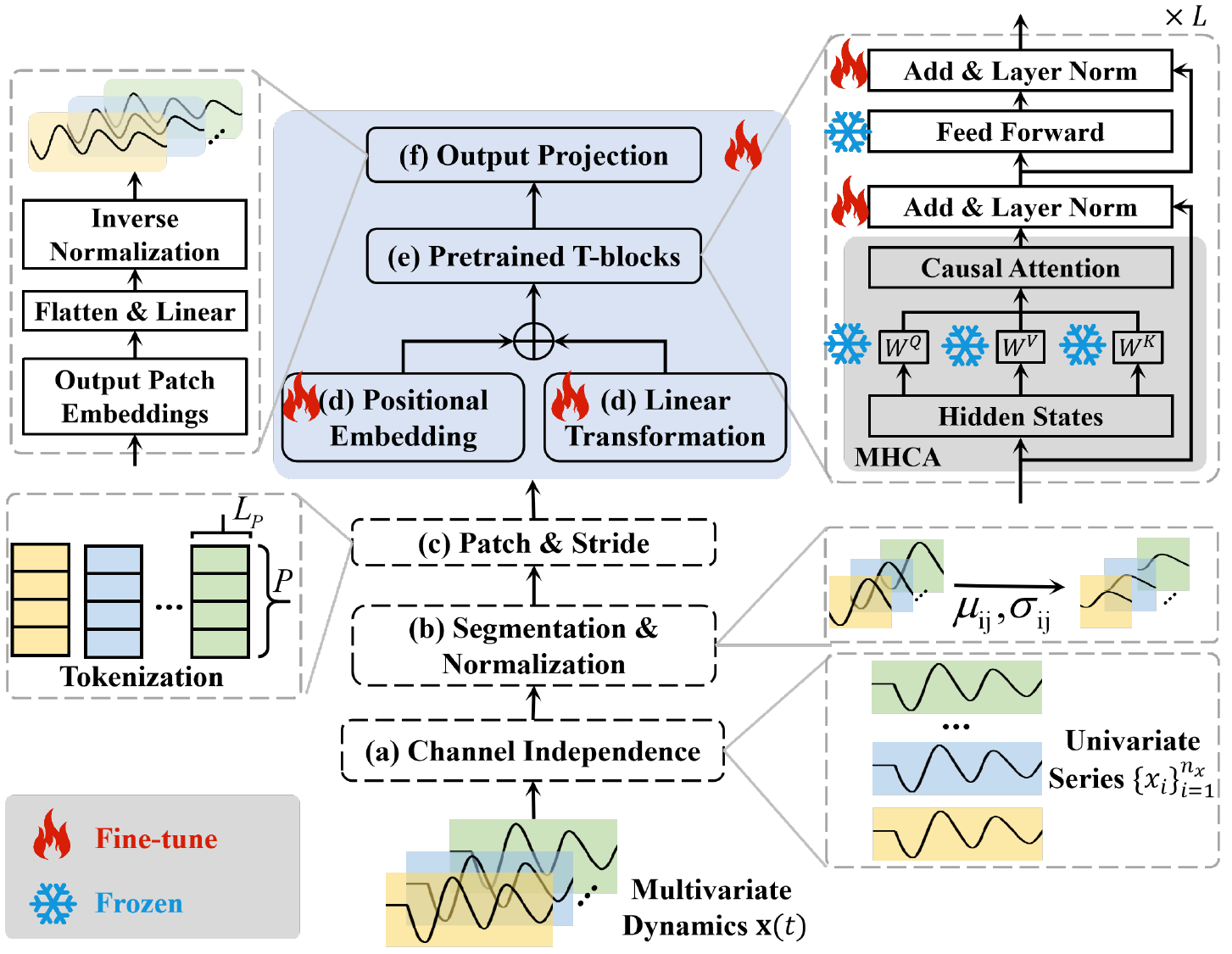}
    \caption{Model structure and pipeline of Uni-TSA.}
    \label{Model Structure of TSA-LLM}
\end{figure}

{\color{black}
This section details the proposed Uni-TSA, covering its data processing paradigm, the GPT-based architecture, and the two-stage fine-tuning scheme. Figure \ref{Model Structure of TSA-LLM} illustrates the overall architecture and data workflow.

\subsection{Data Processing}\label{Data Processing}
\subsubsection{Channel Independence}
{\color{red}
The dynamic state variables of a power system are defined as the multivariate sequences:
\begin{equation}
    \mathbf{x}(t) = [x_1(t), x_2(t), \dots, x_{n_x}(t)]^T,
\end{equation}
where each $x_i(t) \in \mathbb{R}$ corresponds to a distinct state variable channel. Channel independence is formulated at the level of $\mathbf{x}(t)$: rather than treating $\mathbf{x}(t)$ as a fixed-dimensional vector sequence, we decompose it into a set of scalar sequences $\{x_i(t)\}_{i=1}^{n_x}$. Each channel is processed by an identical, shared predictor after applying the same segmentation and temporal-patch tokenization. This channel-wise shared modeling yields a dimension-agnostic input/output interface, and thus eliminates the need to reparameterize the model to match the system-specific value of $n_x$.

For clarity, we use the rotor angle as a concrete example and denote $\delta(t)=[\delta_1(t),\ldots,\delta_{n_g}(t)]^\top$, where $n_g$ is the number of SGs (e.g., $n_g=10$ for the IEEE 39-bus system and $n_g=54$ for the IEEE 118-bus system). Across heterogeneous systems, $n_g$ varies with the network configuration, and so does the dimension of $\delta(t)$. Channel independence addresses this mismatch by mapping $\delta(t)\in\mathbb{R}^{n_g}$ to $\{\delta_i(t)\}_{i=1}^{n_g}$ and applying the same shared predictor to each channel. Consequently, Uni-TSA does not rely on any fixed system-level input/output dimension and can be directly reused across systems with different generator numbers without architecture redefinition.

	Moreover, Uni-TSA keeps the model interface trajectory-centric: the inputs are measured/simulated post-fault trajectories without concatenating explicit topology/parameter tables such as $Y$-bus, line parameters, or generator parameters. In this setting, topology- and parameter-dependent coupling effects are learned
	from the response patterns embedded in the transient trajectories, which preserves a unified
	interface that is convenient to reuse across heterogeneous systems.

	}
\subsubsection{Segmentation and Sample-wise Normalization}
Each univariate sequence $x_i \in \mathbb{R}^T$ is first partitioned into input-target samples via a sliding window of length $L_{\text{sam}} = L_{\text{seq}} + L_{\text{pred}}$ with a unit stride. Each sample contains an input sequence of length $L_{\text{seq}}$ and its corresponding prediction target of length $L_{\text{pred}}$. The input portion, $x_{ij,\text{input}} \in \mathbb{R}^{L_{\text{seq}}}$, is then subjected to {sample-wise normalization} to ensure numerical stability, particularly for high-magnitude unstable trajectories:
\begin{equation}
    x_{ij} = ({x_{ij,\text{input}} - \mu_{ij}})/{\sigma_{ij}} \in \mathbb{R}^{L_{\text{seq}\times1}},\label{standardization}
\end{equation}
where the mean $\mu_{ij}$ and standard deviation $\sigma_{ij}$ are computed from $x_{ij,\text{input}}$ rather than the entire training set. {\color{red}This sample-wise normalization, as presented in \autoref{Model Structure of TSA-LLM}(b), not only manages the extreme values of OOS SGs but also obviates the need for separate normalization statistics for stable/unstable SGs, a common practice in other methods \cite{ye2024use}.} Crucially, our approach eliminates the requirement for an online stability classification that may introduce both processing delays and the risk of error \cite{tan2025bayesian}, thus guaranteeing a more robust and streamlined framework suitable for real-time application.

\subsubsection{Temporal Patch}
The final data processing step is temporal patch, designed to enhance both feature representation and modeling efficiency. Each normalized sample $x_{ij}$ is partitioned into $P$ overlapping patches of length $L_p$ with a stride $S$, yielding a new, more compact representation ${x}^P_{ij} \in \mathbb{R}^{P \times L_p}$ as illustrated in \autoref{Model Structure of TSA-LLM}(c). The number of patches $P$ is given by:
\begin{align}
    P = \left\lfloor ({L_{\text{seq}} - L_p})/{S} \right\rfloor + 1,
\end{align}
where $\lfloor \cdot \rfloor$ is the floor function. This patch-based tokenization enables the model to capture rich, local dynamic patterns (e.g., transient oscillations) that are more informative than isolated time points. Concurrently, by strategically reducing the input sequence length from $L_{\text{seq}}$ to $P$, it is instrumental in mitigating the quadratic complexity ($\mathcal{O}(N^2)$) of the attention mechanism \cite{vaswani2017attention}, thereby ensuring the computational tractability of long-sequence modeling.}

\subsection{Augmented GPT Architecture for TSA}
{\color{black}
{\color{red}
Predicting long-horizon post-fault dynamics from a short observation window is intrinsically ill-posed, because multiple future trajectories may be consistent with the same limited history and small modeling errors can rapidly accumulate over long rollouts \cite{shen2025physics}. To align with the time-evolution nature of post-fault transients, we adopt an iterative autoregressive formulation: the model generates the future trajectory step-by-step, and each prediction is conditioned on the observed history and all previously generated states \cite{zhao2022structure}. This yields a strictly causal factorization of the conditional distribution, which is mathematically consistent with the next-token generation principle in generative LLMs:
}
\begin{align}
	    p\left(\hat{x}_{i j,\text{target}} \mid {x}_{i j,\text{input}}\right)=\prod_{t=L_{\text {scq }}+1}^{L_{\text {sam}}} p\left(x_{i j}(t) \mid x_{i j}(<t)\right).
\end{align}
{\color{red}
Motivated by this alignment, we adopt a Generative Pre-trained Transformer (GPT) backbone \cite{radford2019language} and adapt it for transient trajectory generation. In particular, GPT's causal self-attention naturally enforces temporal causality and provides a strong sequence prior for robust long-horizon autoregressive rollout. The subsequent sections describe the resulting GPT-based architecture of Uni-TSA.
}

\subsubsection{Input Embedding}
The input embedding layer transforms the patch-structured samples ${x}^P_{ij} \in \mathbb{R}^{P \times L_p}$ into the initial hidden state $\mathbf{z}_{i j}^0$. This process is defined as:
\begin{align}
    \mathbf{z}_{i j}^0 = {x}_{i j}^P \cdot \mathbf{W}^p + \mathbf{b}^p + \mathbf{E}_{\text {pos }} \in \mathbb{R}^{P \times d}. \label{input embedding}
\end{align}
In this formulation, each patch in ${x}_{i j}^P$ is first mapped to an embedding of dimension $d$ via a learnable linear projection with weight matrix $\mathbf{W}^p \in \mathbb{R}^{L_p \times d}$ and bias vector $\mathbf{b}^p \in \mathbb{R}^{d}$. Subsequently, a positional encoding matrix, $\mathbf{E}_{\text {pos}} \in \mathbb{R}^{P\times d}$, is added to this projection, which is learned dynamically during fine-tuning to provide the model with essential information about the absolute and relative positions of each patch. This final representation, $\mathbf{z}_{i j}^0$, thus contains both the features of each patch and their sequential context, ready for processing by the transformer layers.

\subsubsection{Transformer Block Architecture}
The backbone of Uni-TSA is composed of a stack of $L$ T-blocks adapted from the GPT model, as illustrated in \autoref{Model Structure of TSA-LLM}(e). Each block performs a transformation that refines an input sequence $\mathbf{z}^{l-1}$ into an output sequence $\mathbf{z}^{l}$:
\begin{align}
    \mathbf{z}^l = \operatorname{Transformer-Block}(\mathbf{z}^{l-1}),~ \text{for } l=1, \dots, L.
\end{align}
This transformation is executed via two primary sub-layers---a multi-head causal self-attention mechanism (MHCA) and a position-wise feed-forward network (FFN)---both embedded within a residual framework that utilizes skip connections and layer normalization.

The MHCA layer containing $h$ parallel causal attention heads, endows the model with the ability to dynamically focus on relevant past information, as illustrated in \autoref{Comparative analysis of attention mechanisms and prediction paradigms.}(b). Within each head, $\mathbf{z}_{ij}^0$ is projected into queries ($\mathbf{Q}$), keys ($\mathbf{K}$), and values ($\mathbf{V}$) using \eqref{QKV} to compute causal attention scores:
\begin{align}
    \text{Cal-atten}(\mathbf{Q}, \mathbf{K}, \mathbf{V}) = \text{softmax}\left(\frac{\mathbf{Q} \mathbf{K}^T}{\sqrt{d_k}} + \mathcal{M}_{\text{causal}}\right) \mathbf{V},
\end{align}
where the lower-triangular mask, $\mathcal{M}_{\text{causal}}$, is formulated as follows. {\color{red}Here, $r$ and $c$ denote the row and column indices (i.e., the query-token position and the key-token position) in the attention-score matrix.}
\begin{align}
    \mathcal{M}_{\text {causal}}(r,c) = 
    \begin{cases}
        0      & \text{if } r \geq c, \\
        -\infty & \text{otherwise}.
    \end{cases}\label{causal_mask}
\end{align}
Pivotal to this operation is the causal mask, $\mathcal{M}_{\text{causal}}$, which nullifies the influence of future tokens by setting their corresponding attention scores to $-\infty$. This architectural constraint is what makes the model inherently suitable for iterative, generative tasks, as it forces the model to generate predictions based solely on available historical data, mimicking online dynamics prediction \cite{liu2024timer}. The outputs of the $h$ heads are subsequently concatenated and linearly transformed by $\mathbf{W}^O$ to yield the multi-head representation as shown in \eqref{Multihead causal transformer}.

\begin{align}
    \text{MHCA}(\mathbf{z}_{ij}^0) = \text{Concat}(\text{Cal-atten}_1, \dots, \text{Cal-atten}_h) \mathbf{W}^O\label{Multihead causal transformer}
\end{align}

The output of the multi-head attention sub-layer, after a residual connection and layer normalization, is then passed through a position-wise FNN, formulated as follows:
\begin{align}
    \begin{aligned}
& \mathbf{h}_{\mathrm{att}}=\operatorname{LayerNorm}\left(\operatorname{MHCA}\left(\textbf{z}_{ij}^0\right)+\textbf{z}_{ij}^0\right), \\
& \textbf{z}_{ij}^1=\operatorname{LayerNorm}\left(\operatorname{FFN}\left(\mathbf{h}_{\mathrm{att}}\right)+\mathbf{h}_{\mathrm{att}}\right),
\end{aligned}
\end{align}
where the FNN module introduces nonlinearity through a two-layer fully connected network:
\begin{align}
    \operatorname{FFN}(\mathbf{h}_{\mathrm{att}})=\operatorname{GELU} (\mathbf{\mathbf{h}_{\mathrm{att}}} \cdot \mathbf{W}_1+\mathbf{b}_1) \cdot\mathbf{W}_2+\mathbf{b}_2.\label{FNN}
\end{align}
In \eqref{FNN}, $\mathbf{W}_1 \in \mathbb{R}^{d \times d_{\mathrm{ff}}}$, $\mathbf{W}_{{2}} \in \mathbb{R}^{d_{\mathrm{ff}} \times d}$
are learnable parameters and $d_{\mathrm{ff}}$ is the hidden dimension. $\operatorname{GELU(\cdot)}$ activation provides a smooth and nonlinear transformation, while the final projection $\textbf{W}_2$ maps features back to the input dimensionality $d$, facilitating the composition of deep transformer architectures through block stacking.

\subsubsection{Output Projection}
\label{sec:output}
The output layer transforms the final latent representation from the transformer, $\mathbf{z}_{ij}^L \in \mathbb{R}^{P \times d}$, into the prediction sequence, as illustrated in \autoref{Model Structure of TSA-LLM}(f). The process begins by flattening the patch-level representations in a single vector ${z}_{ij}^{\text{flat}}=\operatorname{Flatten}({z}_{ij}^L) \in \mathbb{R}^{Pd}$ to consolidate all temporal features. This flattened vector is then mapped to the desired prediction horizon, via a learnable linear projection:
\begin{align}
    {x}_{ij}^{\text{pred}} = {z}_{ij}^{\text{flat}} \cdot \mathbf{W}^{\text{out}} + \mathbf{b}^{\text{out}} \in \mathbb{R}^{L_{\text{pred}}},\label{linear projection}
\end{align}
where $\mathbf{W}^{\text{out}} \in \mathbb{R}^{Pd \times L_{\text{pred}}}$ and $\mathbf{b}^{\text{out}} \in \mathbb{R}^{L_{\text{pred}}}$ are the learnable weights and biases. To restore the prediction to its original physical scale, an inverse normalization operation is applied:
\begin{align}
\hat{x}_{ij,\text{target}}= \sigma_{ij} \cdot {x}_{ij}^{\text{pred}} + \mu_{ij},\label{eq:inversion}
\end{align}
where the mean ($\mu_{ij}$) and standard deviation ($\sigma_{ij}$) are the same values stored from the input sample's normalization step (see \eqref{standardization}). The resulting vector, $\hat{x}_{ij,\text{target}} \in \mathbb{R}^{L_{\text{pred}}}$, {\color{red}represents the final physical prediction for the target sequence of length $L_{\text{pred}}$.}}

\subsection{Two-Stage Fine-tune Strategy}
\subsubsection{Freeze-and-Finetune Strategy}
{\color{black}
To balance adaptation to the TSA task with the preservation of knowledge from pre-trained T-blocks, we propose a parameter-efficient fine-tuning strategy. In this approach, the core T-blocks are kept frozen, while only the parameters of task-specific components, namely the input embedding layer, the layer normalization, and the output projection head, are updated during fine-tuning as illustrated in \autoref{Model Structure of TSA-LLM}. This strategy is designed to preserve the robust, generic feature extraction capabilities that the T-blocks acquire during pre-training. As detailed in \ref{Generic Feature Extraction of Transformer Block}, the T-blocks are inherently optimized to find self- and cross-alignments within input sequences during the pre-training stage, making them powerful, modality-agnostic sequence feature extractors. Freezing the T-blocks leverages this proven capability. The efficacy of this strategy is validated in our case study. The high feature stability and co-direction (detailed in Section \ref{Qualitative Analysis of Pretrained Parameters in LLM}) confirm that the generic features extracted by the pre-trained T-blocks are highly effective for power system dynamics sequences. The following sections detail the proposed two-stage fine-tuning approach, where TeaF trains on sequence segments with ground truth sequences as input, and SchS addresses trajectory-level learning with probabilistically mixed inputs of model predictions and ground truth.}

\begin{algorithm}
\caption{Teacher Forcing Fine-Tuning}
\label{alg:teaf_concise}
\begin{algorithmic}[1]
\STATE \textbf{Input:} Uni-TSA $\mathbf{F}$, Training set $D_{\text{train}}$, Epochs $E_{\text{TeaF}}$, 

Learning rate $\alpha$, Integer $K$
\STATE \textbf{Initialize:} Optimizer for $\mathbf{F}$ with learning rate $\alpha$
\FOR{$\text{k} = 1, 2, \ldots, E_{\text{TeaF}}$}
    \FOR{each sample $(x_{ij,\text{input}}, x_{ij,\text{target}}) \in D_{\text{train}}$}
        \STATE $\hat{x}_{ij,\text{target}} \leftarrow \mathbf{F}(x_{ij,\text{input}})$ 
        \STATE $\mathcal{L}_{\text{TeaF}} \leftarrow \text{MSE}(\hat{x}_{ij,\text{target}}, x_{ij,\text{target}})$
        \STATE Perform back-propagation on $\mathcal{L}_{\text{TeaF}}$ and update $\mathbf{F}$
    \ENDFOR
\ENDFOR
\STATE Initialize empty list $L_{\text{errors}}$
\FOR{each trajectory $x_i \in D_{\text{train}}$}
    \STATE $\hat{x}_i \leftarrow \text{IterativePredict}(\mathbf{F}, x_i)$
    \STATE $e \leftarrow \text{MSE}(\hat{x}_i, x_{i})$, Add $(e, x_i)$ to $L_{\text{errors}}$
\ENDFOR
\STATE Sort $L_{\text{errors}}$ by error in descending order
\STATE $D_{\text{hard}} \leftarrow$ top $K$ trajectories from $L_{\text{errors}}$
\STATE \textbf{Output:} Pre-trained model $\mathbf{F}$, Hard-case set $D_{\text{hard}}$
\end{algorithmic}
\end{algorithm}
\subsubsection{Teacher Forcing Stage}
{\color{black}
During the first TeaF stage, the proposed Uni-TSA, represented by $\operatorname{\mathbf{F}}(\cdot)$, is conditioned on the ground-truth input segments, $x_{ij,\text{input}}$, to generate a prediction for the target sequence (denoted as $\hat{{x}}_{ij,\text{target}} = \operatorname{\mathbf{F}}({x}_{ij,\text{input}})$). The model parameters are optimized by minimizing the mean squared error (MSE) loss, $\mathcal{L}_{\text{TeaF}}$, between the predicted output and the ground-truth target:
\begin{align} 
    \mathcal{L}_{\text{TeaF}} = \frac{1}{L_{\text{pred}}} \sum_{t=1}^{L_{\text{pred}}} \left( \hat{x}_{ij, \text{target}}(t) - x_{ij, \text{target}}(t) \right)^2. 
\end{align}}{\color{black}This TeaF stage, as presented in Algorithm \ref{alg:teaf_concise}, provides a robust model initialization that accelerates convergence in subsequent SchS phases while enabling the model to achieve high fidelity on less complex transient trajectories. It is noteworthy that for numerous existing approaches, their entire training process consists solely of this TeaF paradigm \cite{zhao2022structure, ye2024use, liu2024timer}. However, the iterative prediction process employed in online TSA application, where Uni-TSA recursively conditions on its own outputs as illustrated in \autoref{Comparative analysis of attention mechanisms and prediction paradigms.}(c), presents a training-inference mismatch. The model, trained via TeaF on ground-truth data, suffers from this mismatch and is not equipped to handle its own prediction errors, leading to cumulative error propagation in complex unstable trajectories \cite{lamb2016professor}.}

\subsubsection{Schedule Sampling Phase}
{\color{black}
To mitigate the previously described training-inference mismatch, a scheduled sampling fine-tuning stage is proposed. Following a curriculum learning strategy \cite{lamb2016professor}, SchS focuses on the top-$K$ error-prone trajectories, denoted by $x_i(t)\in D_{\text{hard}}$ for $t\in [1,T]$ identified within $D_{\text{train}}$, as identified in Algorithm \ref{alg:teaf_concise}. {\color{red}This top-$K$ selection, rather than using the full set of trajectories, offers three key advantages: (i) it avoids gradient dilution, since the full set contains many ``easy'' trajectories where the model already performs well after TeaF, whose gradients would dominate and dilute the signal from hard cases that need improvement \cite{yu2025dapoopensourcellmreinforcement}; (ii) it reduces computational cost, as SchS requires full iterative prediction over each trajectory and training on the full set would incur substantially higher per-epoch cost; and (iii) it allocates the limited SchS training budget to the trajectories that most impact online deployment, where cumulative error and training-inference mismatch are most severe.} Critically, by emulating the iterative, self-conditioned prediction of online deployment, this process compels the model to develop robustness against its own error accumulation, thereby improving long-horizon iterative prediction in online TSA tasks.}

{\color{black}
During SchS stage, given an initial input sequence $x_{i,\text{input}}\in \mathbb{R}^{L_{\text{seq}}}$ spanning time steps $[1:L_{\text{seq}}]$ from trajectory $x_i$, the model  first generates a prediction for the next time window $[L_{\text{seq}}+1,L_{\text{seq}}+L_{\text{pred}}]$:
\begin{align}
\hat{\mathbf{x}}_{i, \text{target}}=\mathbf{F}\left(x_{i, \text{input}}\right)\in \mathbb{R}^{L_{\text{pred}}}.
\end{align}
To generate the subsequent prediction, a new input sequence, $x_{i,\text{input}}^{\text{next}}$, is constructed by sliding the window forward. The core of the SchS strategy lies in how this new input is assembled. It consists of a known segment from the original ground-truth trajectory ($x_{i,\text{orig}}$) and a "mixed" segment ($x_{i,\text{mixed}}$), which is probabilistically sampled from either the ground-truth target ($x_{i, \text {target}}$) or the model's own previous prediction ($\hat{x}_{i, \text {target}}$):
\begin{align}
    x_{i, \text {mixed}} = 
    \begin{cases}
        x_{i, \text {target}}, & \text{with probability } \varepsilon_k \\ 
        \hat{x}_{i, \text {target}}, & \text{with probability } 1-\varepsilon_k
    \end{cases}\label{mixed_input}
\end{align}
where $\varepsilon_k$ is the sampling rate for the current epoch $k$. The new input for the model is then a concatenation of these segments:
\begin{align}
    x_{i, \text {input}}^{\text{next}} = [x_{i, \text {orig}}, x_{i, \text {mixed}}]. \label{time segment combination}
\end{align}
This probabilistic mixing is governed by a sampling rate $\varepsilon_k$ that follows a linear decay schedule across training epochs. This gradually transitions the model from teacher forcing ($\varepsilon_k=1$) to a fully iterative mode ($\varepsilon_k=0$):
\begin{align}
    \varepsilon_k= \begin{cases}1, & \text { if } k<E_{\text {start}} \\ 1-\frac{k-E_{\text {start}}}{E_{\text {max}}-E_{\text {start}}}, & \text { otherwise }\end{cases}\label{sample rate}
\end{align}
where $E_{\text{start}}$ is the epoch at which decay begins, and $E_{\text{max}}$ is the maximum training epoch in SchS stage. The fine-tuning loss, $\mathcal{L}_{\mathrm{SchS}}$, aggregates the MSE across the predicted trajectory sequence $\hat{x}_{i}(t)$:
\begin{align}
    \mathcal{L}_{\mathrm{SchS}}=\frac{1}{(T-L_{\mathrm{seq}})} \sum_{t=L_{\mathrm{seq}}+1}^T\left(\hat{x}_{i}(t)-x_{i}(t)\right)^2.
\end{align}
The complete SchS procedure is detailed in Algorithm \ref{SchS stage}.
}

\begin{algorithm}[H]
\caption{Scheduled Sampling Fine-Tuning}
\label{SchS stage}
\begin{algorithmic}[1]
\STATE \textbf{Input:} Pre-trained model $\mathbf{F}$, Hard-case set $D_{\text{hard}}$, Epochs $E_{\text{max}}$, Scheduled epoch $E_{\text{start}}$, Learning rate $\alpha$
\STATE \textbf{Initialize:} Optimizer for $\mathbf{F}$ with learning rate $\alpha$
\FOR{$k = 1, 2, \ldots, E_{\text{max}}$}
    \STATE Calculate sampling probability $\varepsilon_k$ using Eq. \eqref{sample rate}
    \FOR{each trajectory $x_{i} \in D_{\text{hard}}$}
        \STATE \textbf{Initialize:} Total loss $\mathcal{L}_{\text{SchS}} \leftarrow 0$
        \STATE $x_{i, \text{input}} \leftarrow x_{i}(1:L_{\text{seq}})$
        
        \FOR{$j = 1, 2, \ldots, \lfloor(T - L_{\text{seq}})/L_{\text{pred}}\rfloor$} 
            \STATE $\hat{x}_{i, \text{target}} \leftarrow \mathbf{F}(x_{i, \text{input}})$ 
            \STATE $\mathcal{L}_{\text{SchS}} \leftarrow \mathcal{L}_{\text{SchS}} + \text{MSE}(\hat{x}_{i, \text{target}}, x_{i,\text{target}})$
            \STATE $x_{i, \text {input}}^{\text{next}} \leftarrow$ Construct next input with $x_{i,\text{input}}$, $\hat{x}_{i,\text{target}}$, 
            
            ${x}_{i,\text{target}}$, and $\varepsilon_k$ based on \eqref{mixed_input} and \eqref{time segment combination}
            \STATE $x_{i, \text{input}} \leftarrow x_{i, \text {input}}^{\text{next}}$
        \ENDFOR
        \STATE Perform back-propagation on $\mathcal{L}_{\text{SchS}}$
        \STATE Update model $\mathbf{F}$ parameters
    \ENDFOR
\ENDFOR
\STATE \textbf{Output:} Final fine-tuned Uni-TSA 
\end{algorithmic}
\end{algorithm}

\subsection{Online  Application}\label{Real World Application}


{\color{black}
While channel independence processes each of the $n_x$ state variables separately in the offline fine-tuning stage, we mitigate potential computational overhead in online applications by treating the $n_x$ channels as a unified mini-batch to exploit GPU parallelization and concurrent predict of all state variables. The online prediction begins with an initial multivariate observation from PMUs, denoted as $\textbf{x}^{(0)}_{\text{input}}\in \mathbb{R}^{n_\textbf{x}\times L_{\text{seq}}}$.
This multi-channel input is decomposed into $n_x$ independent univariate patch sequences $\mathbf{x}_{\text{patch}}^{(0)}\in \mathbb{R}^{P\times L_P}$ as described in \autoref{Data Processing}, which are then stacked along the batch dimension via $\operatorname{Batch}(\cdot)$, forming $\textbf{x}^{(0)}_{\text{batch}} \in \mathbb{R}^{n_\textbf{x} \times P \times L_p}$. Uni-TSA then predicts the target sequence of all state variables $\hat{\textbf{x}}^{(0)}_{\text{target}} \in \mathbb{R}^{n_\textbf{x} \times L_{\text{pred}} \times 1}$. Then, iterative prediction for each step $j$ is performed sequentially as follows: 
\begin{align}
    &j\in[1, \left\lfloor\tfrac{T-L_{\text{seq}}}{L_{\text {pred}}}\right\rfloor]:~\begin{cases}
\mathbf{x}^{(j)}_{\text{input}} = \mathbf{x}^{(j-1)}_{\text{input}} \oplus \hat{\mathbf{x}}_{\text{target}}^{(j-1)}, \\
\mathbf{x}^{(j)}_{\text{patch}}=\operatorname{Preprocess}(\mathbf{x}^{(j)}_{\text{input}}),
\\
\mathbf{x}^{(j)}_{\text{batch}}=\text{Batch}(\mathbf{x}^{(j)}_{\text{patch}}),\\
\hat{\mathbf{x}}_{\text {target}}^{(j)}=\mathbf{F}\left(\mathbf{x}^{(j)}_{\text{batch}}\right),
\end{cases}\label{iterative prediction}
\end{align}
where $\oplus$ denotes the sliding window update mechanism, which discards the oldest input segment and appends the latest prediction, and $\operatorname{Preprocess}(\cdot)$ follows the data processing pipeline in \autoref{Data Processing}. This fully iterative online prediction aligns with the SchS strategy where the sampling probability $\epsilon_k$fixed at 0, forcing the model to rely entirely on its own generated trajectory.}

{\color{blue}
\subsection{Data Flow and Dimensional Transformations}\label{sec:data_flow}

To provide a transparent and reproducible blackprint of the proposed architecture, this subsection systematically summarizes the exact data dimensions and matrix shapes along the data flow. As illustrated in \autoref{fig:dim_change}, the structural alignment of these dimensions between the training and online application phases serves as the core mathematical basis that enables the model's cross-system scalability.

\begin{figure}[htbp]
    \centering
    \includegraphics[width=0.9\linewidth]{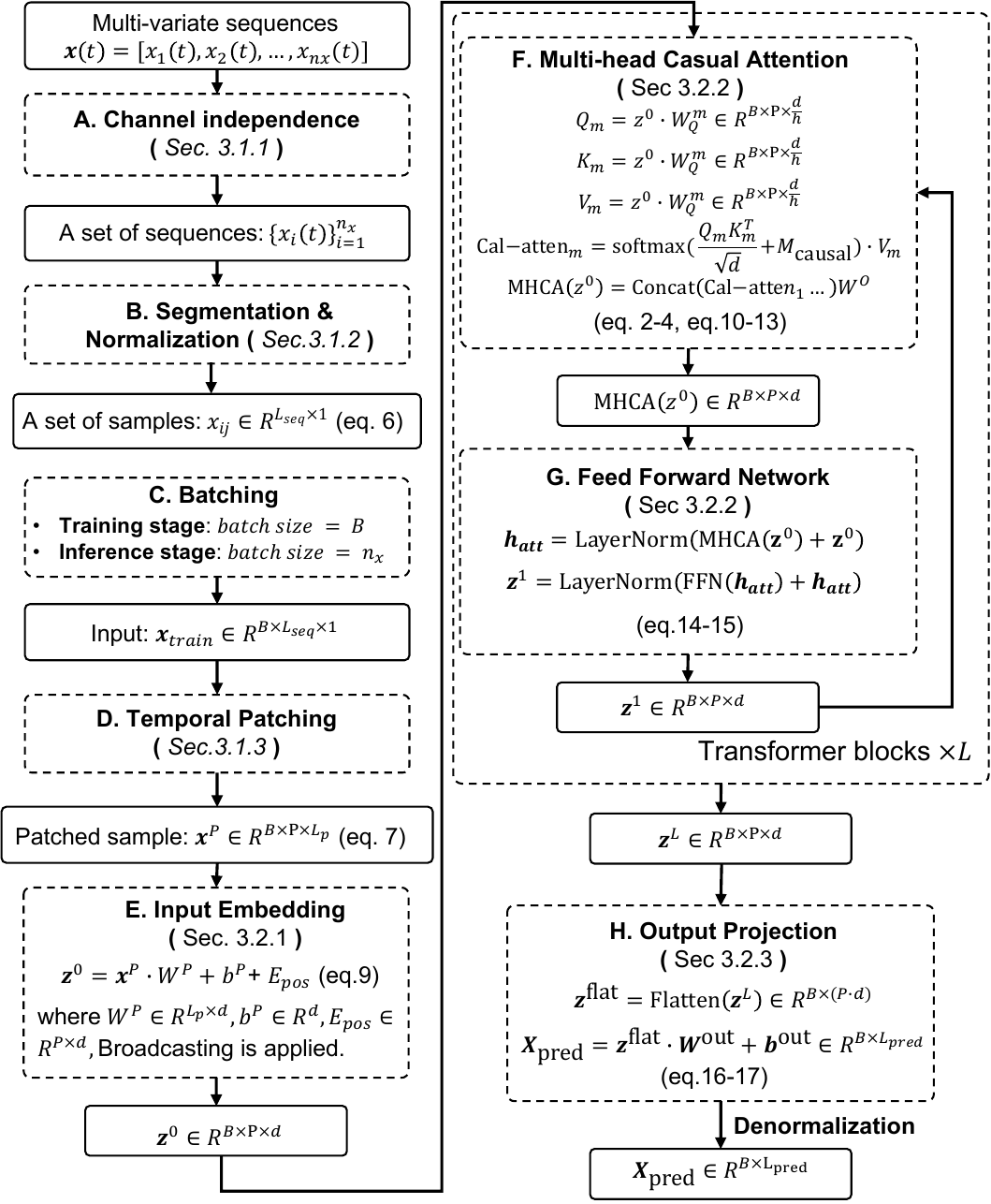}
    \caption{Detailed illustration of the data flow and dimensional transformations within the Uni-TSA framework. The solid boxes represent the sequential processing modules during forward propagation, while the dashed boxes explicitly denote the tensor shapes of the intermediate features output by each corresponding module. For reproducibility, the specific mathematical operations and their corresponding section/equation indices in the manuscript are annotated alongside the data flow.}
    \label{fig:dim_change}
\end{figure}

During the training phase, the channel independence (CI) mechanism treats the trajectories of different state variables as individual, independent samples. Let $B$ denote the training batch size, $L_{\text{seq}}$ the input sequence length, and $L_{\text{pred}}$ the prediction horizon length. The detailed data transformations are as follows:
\begin{enumerate}
    \item \textbf{Input Formulation}: A batch of $B$ independent, normalized univariate input segments, denoted as $x_{ij} \in \mathbb{R}^{L_{\text{seq}}}$ in \eqref{standardization}, is sampled. This forms the initial input tensor $\mathbf{x}_{\text{train}} \in \mathbb{R}^{B \times L_{\text{seq}} \times 1}$.
    \item \textbf{Temporal Patching}: As detailed in Section~\ref{Data Processing}, the continuous sequence $x_{ij}$ is partitioned into overlapping patches of length $L_p$. The sequence representation for the batch thus becomes $\mathbf{x}^P \in \mathbb{R}^{B \times P \times L_p}$, corresponding to a batch of the patch-structured samples ${x}^P_{ij}$.
      \item \textbf{Embedding Layer}: Through a learnable linear projection, each $L_p$-dimensional patch is mapped into the hidden model dimension $d$. After adding the positional encoding $\mathbf{E}_{\text{pos}}$ as formulated in \eqref{input embedding} (which is broadcasted to every sample in the batch), the embedded representation for the batch attains the dimension $\mathbf{z}^0 \in \mathbb{R}^{B \times P \times d}$, which concatenates the instance-level embeddings $\mathbf{z}_{ij}^0$.
    \item \textbf{Transformer Blocks Processing}: The embedded tensor $\mathbf{z}^0$ is fed into the stacked Transformer blocks. Within each causal attention head, the queries ($\mathbf{Q}$), keys ($\mathbf{K}$), and values ($\mathbf{V}$) computed using \eqref{QKV} are projected into the shape $\mathbb{R}^{B \times P \times (d/h)}$, where $h$ is the total number of attention heads. The intermediate causal attention score matrix, regulated by the lower-triangular mask $\mathcal{M}_{\text{causal}}$ in \eqref{causal_mask}, has a shape of $\mathbb{R}^{B \times P \times P}$. After executing multi-head concatenation \eqref{Multihead causal transformer}, residual addition, and processing through the position-wise feed-forward networks \eqref{FNN}, each Transformer block preserves the hidden shape. Accordingly, the final output representation of the Transformer stack remains $\mathbf{z}^L \in \mathbb{R}^{B \times P \times d}$, which corresponds to the batch equivalent of $\mathbf{z}_{ij}^L$.
    \item \textbf{Output Projection}: To decode the future sequence, the hidden representation $\mathbf{z}^L$ is flattened to $\mathbf{z}^{\text{flat}} \in \mathbb{R}^{B \times (P \cdot d)}$. A subsequent linear prediction head is applied based on \eqref{linear projection}, producing the final continuous prediction tensor $\mathbf{X}_{\text{pred}} \in \mathbb{R}^{B \times L_{\text{pred}}}$, encompassing the individual predictions ${x}_{ij}^{\text{pred}}$.
\end{enumerate}

The data flow in the online application phase is structurally identical to that in the training phase, with the fundamental distinction being that the algorithmic batch size dimension $B$ is physically replaced by the number of system state variables $n_x$. During online deployment, the model concurrently predicts the future trajectories for all $n_x$ variables of a given specific power system. By substituting $B$ with $n_x$, the initial multivariate online observation is naturally stacked into $\mathbf{x}^{(0)}_{\text{batch}} \in \mathbb{R}^{n_x \times L_{\text{seq}} \times 1}$. Following the identical dimensional transformations, the parallelized tensor sequentially evolves to $\mathbb{R}^{n_x \times P \times L_p}$ after patching, $\mathbb{R}^{n_x \times P \times d}$ after embedding and Transformer-block updates, and finally yields the concurrent predictions $\hat{\mathbf{x}}^{(0)}_{\text{target}} \in \mathbb{R}^{n_x \times L_{\text{pred}}}$ for all $n_x$ variables simultaneously.

Through this carefully aligned dimensional design, the model processes one individual channel as one batch sample during training, while during online inference, the $n_x$ physical channels are processed in parallel by treating $n_x$ directly as the batch size. Consequently, varying the scale of the target power system only alters the batch dimension $B$ fed into the GPU rather than any parameter shapes of the model weights itself, inherently guaranteeing the model's architectural zero-shot transferability across heterogeneous systems.
}

\subsection{Performance Indices Evaluation}\label{Performance Indices Evaluation}
{\color{black}
To provide a comprehensive analysis, the model's performance is evaluated on three distinct stability situations. The stable (S) situation comprises trajectories exclusively from stable OCs, while the unstable (U) situation contains only those from unstable OCs. A third, hybrid (H) situation is composed of a combination of both stable and unstable trajectories to assess overall generalization.
The performance metrics ($MAE$, $MSE$) for each category $\mathrm{C} \in \{\mathrm{S}, \mathrm{U}, \mathrm{H}\}$ are formally defined by averaging the error over all trajectories in the corresponding stability situations:
\begin{equation}
\begin{aligned}
{MAE}_{\mathrm{C}} & =\sum_{x \in \mathcal{D}_{\mathrm{C}}} \sum_{i=1}^{n_x} \sum_{t=L_{\text {seq }}+1}^T\frac{\left|\hat{x}_i(t)-x_i(t)\right|}{|{D}_{\mathrm{C}}|\cdot n_x\cdot\left(T-L_{\text {seq}}\right)}, \\
{MSE}_{\mathrm{C}} & =\sum_{x \in \mathcal{D}_{\mathrm{C}}} \sum_{i=1}^{n_x} \sum_{t=L_{\text {seq }}+1}^T\frac{\left(\hat{x}_i(t)-x_i(t)\right)^2}{|{D}_{\mathrm{C}}|\cdot n_x\cdot\left(T-L_{\text {seq}}\right)},
\end{aligned}
\end{equation}
where ${D}_{\mathrm{C}}$ is the test set for category C, $|{D}_{\mathrm{C}}|$ is the number of trajectories within it, and $T_{\text{pred}} = T-L_{\text{seq}}$ is the prediction length. The terms $\hat{x}_{i}(t)$ and $x_{i}(t)$ denote the predicted and ground-truth values, respectively, for the $i$-th dynamic trajectories at time $t$. This categorized analysis allows us to not only quantify but also explain the performance disparities observed in its handling of stable versus unstable system dynamics.
}

\section{Case study}
The proposed Uni-TSA framework is rigorously evaluated on the New England 39-bus system \cite{zhao2022structure} and the modified Iceland 189-bus system \cite{cai2020data}. On the New England 39-bus system, we assess the framework's zero-shot generalization to mixed-stability OCs and unseen faults and conduct a representation analysis of its unstable OC handling mechanism. The Iceland 189-bus system is used to validate the model's few-shot scalability on a more complex, heterogeneous grid. The evaluation is completed with analyses of training/inference costs, an ablation study of model components, and a final qualitative demonstration of the predictive enhancements gained from using frozen T-blocks.

\subsection{Data Generation and Model Settings}
\subsubsection{Test System and Simulation Settings}
{\color{black}
Two standard test systems were applied: the New England 39-bus system (10 SGs, 34 lines) and the modified Iceland 189-bus system (33 SGs, 206 lines). The latter was adapted for 20\% RES penetration by replacing two SGs with wind units whose speeds were sampled from a Weibull distribution \cite{lu2024advanced, shu2015investigation}. Simulations include $N-1$, $N-2$, and $N-3$ faults for the New England 39-bus system and $N-1$ faults for the Iceland 189-bus system. For each contingency, simulations are conducted as follows: 1) All system loads were randomly varied between 70\% and 130\% of their nominal values; 2) An optimal power flow was solved using MATPOWER to establish the pre-fault OCs; 3) All SGs were represented by the fourth-order model; 4) Each contingency was simulated by applying the specified number of concurrent three-phase-to-ground faults to random lines at $t=1.0~s$ and cleared after a random duration of up to 0.3 s; 5) The post-fault dynamic response was simulated for 10 seconds with a 0.02 s time step by Power System Analysis Toolbox.}

\subsubsection{Dataset Organization}
{\color{black}
For fine-tuning and validation, a primary dataset of 9,000 trajectories (rotor angles and speeds) was generated from $N-1$ contingencies on the 39-bus system, covering a mix of stable and unstable trajectories. This was partitioned into 8,000 training ($D_{\text{train}}$) and 1,000 validation ($D_{\text{val}}$) trajectories. Each trajectory was then segmented via a sliding window ($L_{\text{seq}}=65, L_{\text{pred}}=1$) to yield approximately $1.8 \times 10^7$ training and $2 \times 10^6$ validation samples. Additionally, trajectories from the 189-bus system were generated to evaluate cross-system few-shot adaptation on a more complex, heterogeneous grid. Model performance was evaluated against four distinct test sets, each designed to assess a specific aspect of universality. The details of these datasets are summarized in \autoref{Test Set Overview}. The subsequent analysis primarily presents rotor angle predictions and results for other variables like rotor speeds are analogous and detailed in \autoref{angular speed test}.
{\color{red}
For the heterogeneous-system adaptation study on the 189-bus system, we generated a target-system corpus of 9,900 trajectories under $N\!-\!1$ contingencies and randomly split it into $D_{\text{train}}^{189\text{bus}}$ (6,600 trajectories) and $D_{\text{test}}^{189\text{bus}}$ (3,300 trajectories). Unless otherwise stated, the \emph{few-shot} setting uses 5\% of $D_{\text{train}}^{189\text{bus}}$ (i.e., 330 trajectories) for adaptation and reports the results on $D_{\text{test}}^{189\text{bus}}$. The few-shot subset is sampled at the trajectory level (rather than at the sliding-window sample level) to avoid information leakage across overlapping windows, and the same segmentation pipeline is then applied to construct supervised input--target pairs.
To further illustrate the data-efficiency trend of cross-system adaptation, we additionally report results under increasing target-data ratios on other heterogeneous benchmark systems (\autoref{sec:scaling_68_118}).
}

\begin{table}[H]
  \setlength{\tabcolsep}{3pt} 
  \renewcommand{\arraystretch}{1.5} 
  \centering
  \caption{Test Sets Overview}

    \begin{tabular}{cccccc}
    \hline
    Test Set & Tested Universality & Test Mode & System & Fault & Trajectories \bigstrut\\
    \hline
    $D_{\text{test}}^{N-1}$ & Mixed OCs & Zero-shot & 39-bus & $N-1$ & 1000\bigstrut[t]\\
    $D_{\text{test}}^{N-2}$ & Unseen faults & Zero-shot & 39-bus &$N-2$ & 1,000
    \\
    $D_{\text{test}}^{N-3}$ & Unseen faults & Zero-shot & 39-bus &$N-3$ & 1,000 \\
    $D_{\text{test}}^{189\text{bus}}$ & \makecell{Heterogeneous\\system} & Few-shot & 189-bus & $N-1$ & 3,300 \bigstrut[b]\\
    \hline
    \end{tabular}%
  \label{Test Set Overview}%
\end{table}%
}

\subsubsection{Baseline Selection Rationale}
{\color{black}
The selection of baseline methods for comparison is guided by three key principles: (1) {representativeness of state-of-the-art TSA methods}, (2) {architectural diversity to isolate the contribution of key design choices}, and (3) {reproducibility and fair comparison}. 

First, we include the DNR framework \cite{zhao2022structure} {\color{red}(143K parameters)} and LSTM-based approach \cite{ye2024use} {\color{red}(34K parameters)} as they represent the current state-of-the-art in TSA trajectory prediction. DNR is selected because it is the most recent unified framework that handles both stable and unstable OCs within a single model, making it the strongest existing baseline for universality evaluation. The LSTM baseline \cite{ye2024use} is included as it represents the classical recurrent architecture widely adopted in TSA trajectory prediction, enabling us to assess the benefits of transformer-based architectures over traditional RNNs. Notably, LSTM has been the most frequently used baseline for comparison in recent TSA trajectory prediction studies \cite{zhao2022structure, tan2025bayesian}, further justifying its inclusion. Second, we include an Encoder-only Transformer (ENC) baseline {\color{red}(9.6M parameters)} to isolate the contribution of causal attention mechanisms. The ENC baseline employs bidirectional attention, contrasting with our causal attention mechanism, thereby allowing us to demonstrate the importance of temporal causality in iterative prediction tasks.

Several methods referenced in \autoref{Universality Comparison of the Existing and Proposed Methods} (PINN \cite{misyris2020physics}, PFNN \cite{shen2025physics}, FNO \cite{cui2023frequency}, SVDKR \cite{tan2025bayesian}) are excluded from quantitative comparison due to fundamental methodological incompatibilities: (1) PINN lacks explicit design considerations for the three universality dimensions addressed in this work—namely, mixed stable/unstable operating conditions, generalization to unseen fault scenarios, and cross-system transferability—rendering it unsuitable for universality evaluation; (2) FNO operates in the frequency domain and necessitates distinct data preprocessing pipelines, introducing confounding factors that preclude equitable comparison with time-domain approaches; (3) SVDKR adopts a probabilistic prediction paradigm, which fundamentally differs from deterministic trajectory forecasting and impedes direct quantitative assessment against deterministic baselines. Furthermore, preliminary evaluation of PFNN \cite{shen2025physics} revealed performance characteristics comparable to the classical LSTM baseline under universal scenarios, thereby providing limited additional insight. Consequently, our comparative analysis focuses on LSTM, DNR, and ENC, which collectively encompass the requisite architectural diversity and constitute the most pertinent baselines for evaluating universality in TSA trajectory prediction.
}

\subsubsection{Model Implementation Details}
{\color{black}
The proposed Uni-TSA, based on the GPT architecture (124M parameters, 12 layers, 12 heads, 768-dim embedding), was fine-tuned for up to 10 epochs using a two-stage TeaF and SchS scheme. Training utilized the Adam optimizer with a $1 \times 10^{-4}$ initial learning rate managed by a cosine annealing scheduler and an early stopping criterion. To evaluate its performance, Uni-TSA was benchmarked against three baseline models. The first two are the DNR framework from \cite{zhao2022structure} and a standard LSTM network as presented in \cite{ye2024use}, for which we employed the hyperparameter settings as reported in their respective publications. The third baseline is an Encoder-only Transformer (ENC), constructed based on \cite{vaswani2017attention} to highlight performance differences arising from different attention mechanisms. This ENC model employs the bidirectional attention mechanism and is configured with 3 transformer layers, 6 bidirectional attention heads, and a 128-dim embedding. {\color{red}The parameter counts are 124M (total) and 858K (trainable, 0.69\% of the total) for Uni-TSA, 34K for LSTM \cite{ye2024use}, 143K for DNR \cite{zhao2022structure}, and 9.6M for ENC.} All baselines were implemented for multivariate dynamics prediction in PyTorch and trained on NVIDIA A100 GPU.
}

\subsection{Evaluation on Mixed Stability OCs}\label{State Prediction Capacity Comparison}



\begin{figure}
    \centering
\subfloat[Actual $\delta$ in stable OC]{\includegraphics[width=0.47\linewidth]{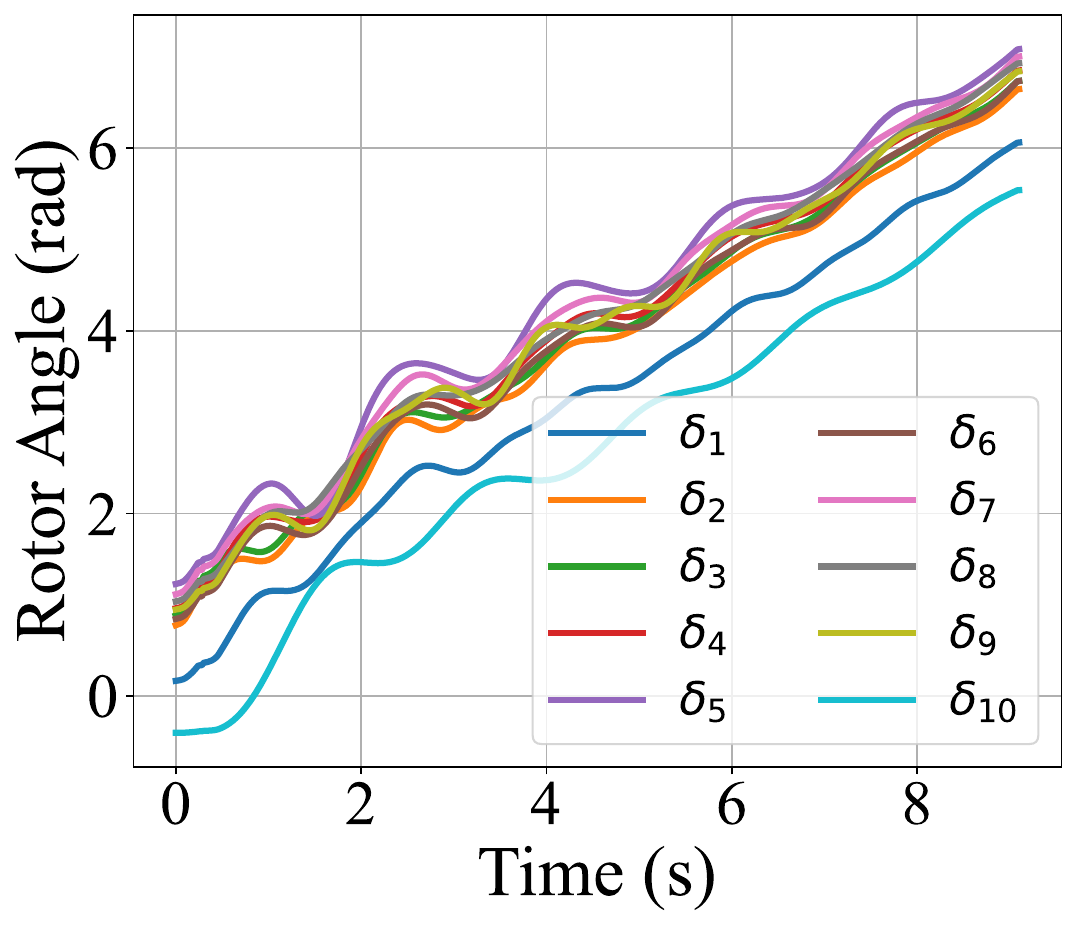}}
\subfloat[Actual $\delta$ in unstable OC]{\includegraphics[width=0.47\linewidth]{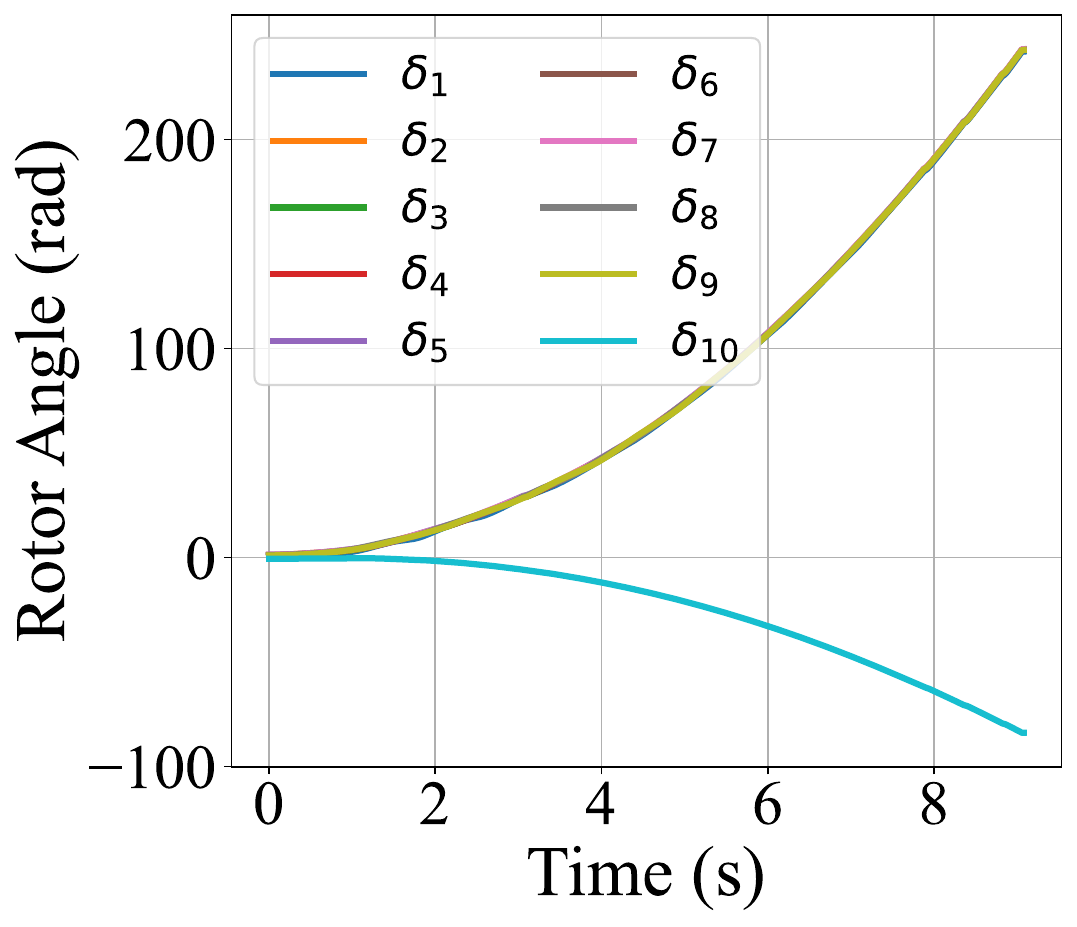}}\\
    \subfloat[Predicted $\delta$ in stable OC]{\includegraphics[width=0.47\linewidth]{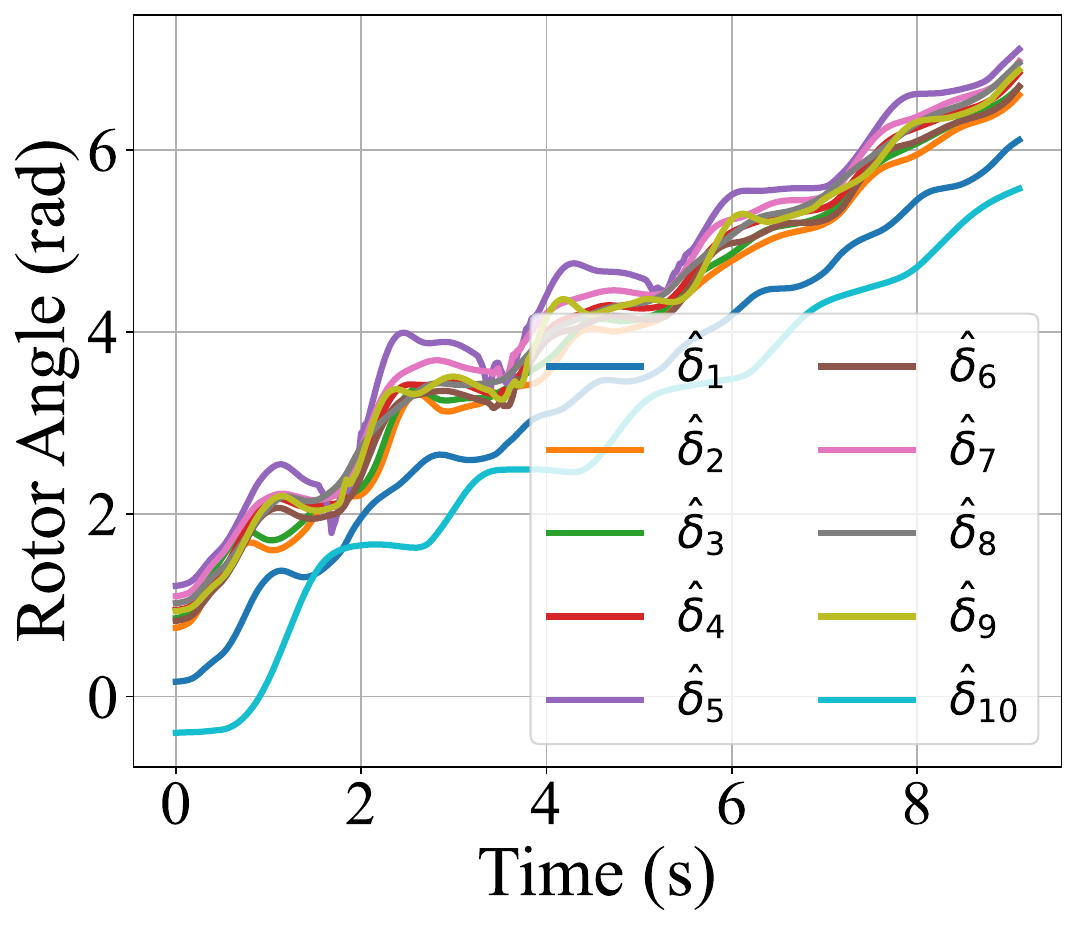}}
        \subfloat[Predicted $\delta$ in unstable OC]{\includegraphics[width=0.47\linewidth]{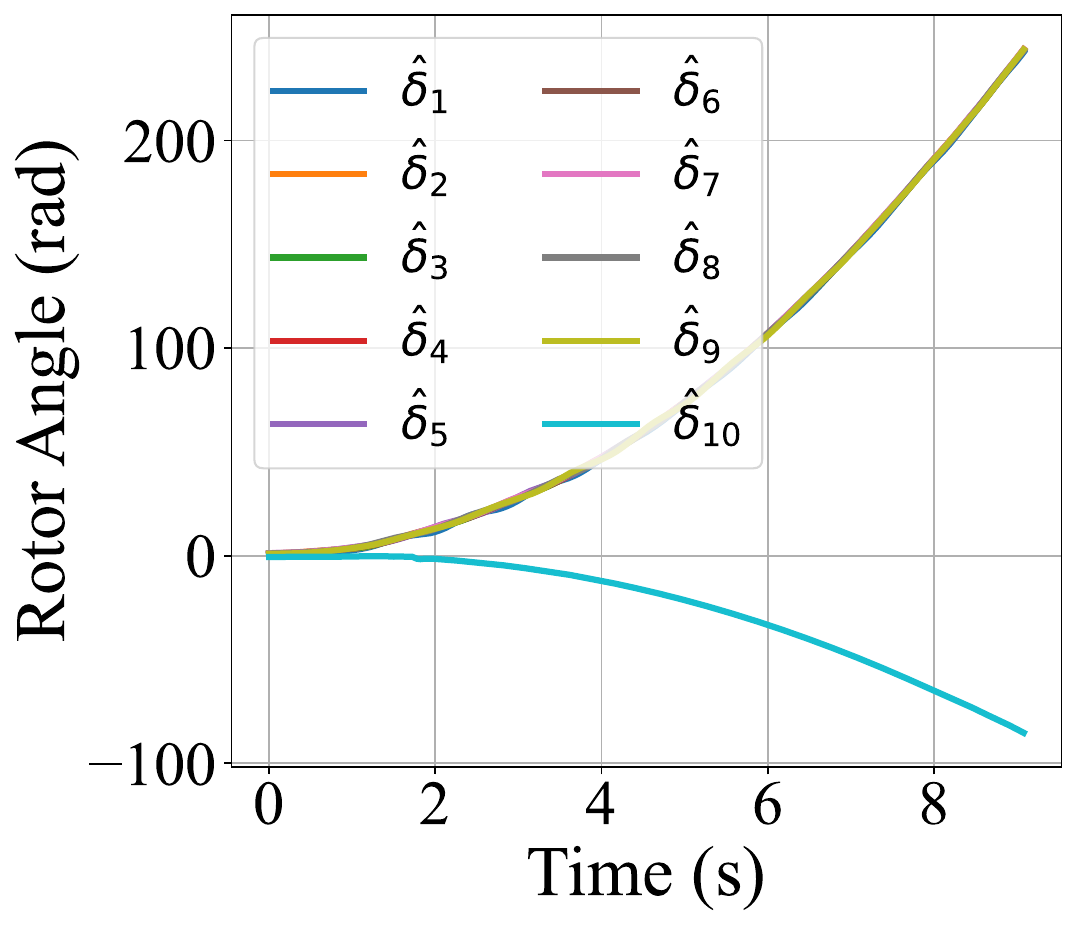}}
    \caption{New England 39-bus system: The evaluation of rotor angle under the stable and unstable OC. The predicted $\delta$ by Uni-TSA ((c) and (d)) demonstrates close alignment with ground truth ((a) and (b)).}
    \label{New England 39-bus system: The evaluation of rotor angle under the stable and unstable condition.}
\end{figure}

{\color{black}
\subsubsection{Performance Comparison}
This section evaluates the model's robustness against mixed stable or unstable OCs on the $D_{\text{test}}^{N-1}$, which is collected in $\{\text{S},\text{U},\text{H}\}$ situations as discussed in \autoref{Performance Indices Evaluation}. The quantitative results are summarized in \autoref{New England 39-Bus System: State Prediction Capacity Comparison}, where bold and underlined values denote the best and second-best performance, respectively. As shown, Uni-TSA substantially outperforms all baseline models across nearly all metrics, with the performance gap being most pronounced in challenging unstable and hybrid cases. While traditional methods like DNR and LSTM struggle significantly in these scenarios with DNR's $MSE_U$ value even exceeding 100, Uni-TSA achieves remarkable error reductions. Specifically, Uni-TSA reduces $MAE_U$, $MSE_U$, $MAE_H$, and $MSE_H$ by at least 84.49\%, 97.27\%, 83.49\%, and 97.68\%, respectively, compared to baselines. These quantitative findings are further corroborated by visualization of the rotor angle trajectories predicted by Uni-TSA in \autoref{New England 39-bus system: The evaluation of rotor angle under the stable and unstable condition.}, which show a close tracking of the ground truth. {\color{red}For improved clarity, \autoref{Representative trajectory comparison for 39-bus SG 10} further presents representative trajectory comparisons for SG 10 with predictions and ground truths plotted on the same axes.} This significant disparity in performance can be attributed to fundamental architectural differences. Traditional RNN-based models (LSTM and DNR) suffer from vanishing gradient issue and exhibit constrained model capacity, which fundamentally limits their ability to capture complex nonlinear dynamics in unstable trajectory prediction. In contrast, the transformer architecture, utilized by both our proposed Uni-TSA and the ENC baseline, is far more effective at capturing long-range temporal dependencies, resulting in demonstrably superior accuracy in these challenging scenarios.
}

\subsubsection{Explaining Robust Prediction in Unstable Scenarios} 
\begin{figure}
    \centering
    \includegraphics[width=0.9\linewidth]{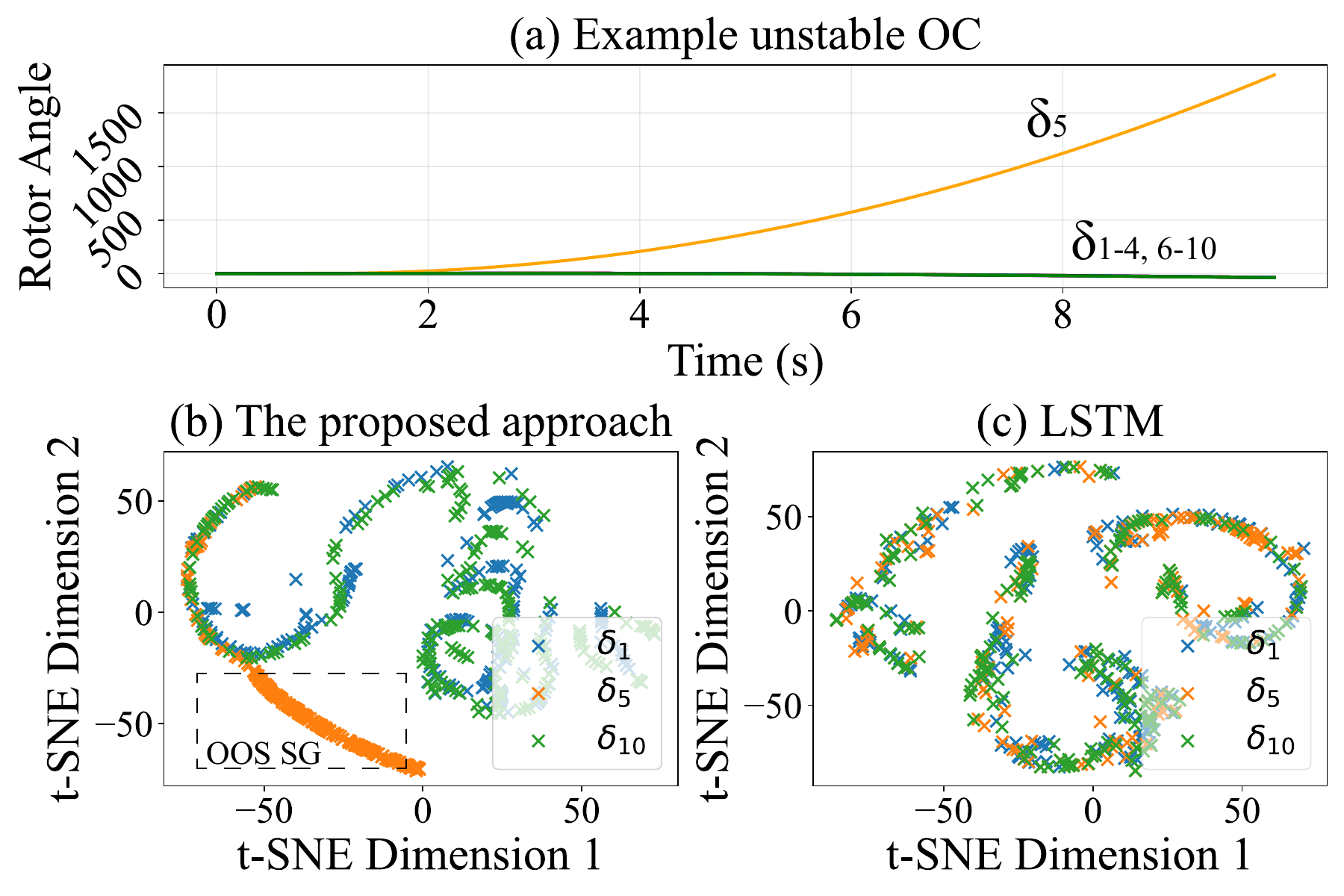}
    \caption{The t-SNE visualization of stable/unstable sample feature maps for the proposed Uni-TSA and LSTM.}
    \label{The t-SNE visualization of sample feature maps for the proposed approach and LSTM.}
\end{figure}

{\color{black}
To further investigate the source of Uni-TSA's robustness, we conducted an interpretive analysis focused on its feature representations of OOS dynamics, which are critical in unstable scenarios. We selected a typical unstable case from the 39-bus system where SG $\delta_5$ loses synchronism (\autoref{The t-SNE visualization of sample feature maps for the proposed approach and LSTM.}(a)). Final-layer feature representations for the OOS SG ($\delta_5$) and two stable SGs ($\delta_1, \delta_{10}$) were then extracted and projected into 2D space from both Uni-TSA and an identically trained channel-independent LSTM model for visualization with t-SNE \cite{zhou2023one}. The t-SNE visualizations reveal a stark contrast between the two architectures. For Uni-TSA (\autoref{The t-SNE visualization of sample feature maps for the proposed approach and LSTM.}(b)), features from $\delta_5$ form a distinct and clearly separable cluster, indicating effective learning of instability characteristics. However, the LSTM model exhibits substantial mixing of these representations (\autoref{The t-SNE visualization of sample feature maps for the proposed approach and LSTM.}(c)). This demonstrates the inherent difficulty of RNN architectures in distinguishing the unique behavior of OOS generators, which correlates with their lower prediction accuracy and underscores the superior feature learning of the proposed attention-based architecture.}

\begin{table}
  \centering
  \setlength{\tabcolsep}{3pt} 
  \caption{New England 39-Bus System: Evaluation Results on Mixed Stability OCs}
    \begin{tabular}{ccccccc}
    \hline
    Model & $MAE_S$ & $MSE_S$ & $MAE_U$ & $MSE_U$ & $MAE_H$ & $MSE_H$ \bigstrut\\
    \hline
    LSTM  & {0.120}  & 0.282  & 2.675  & 19.269  & 0.946  & 5.932  \bigstrut[t]\\
    DNR   & 0.128  & 0.200  & 7.073  & 133.614  & 1.934  & 35.367  \\
    ENC & \underline{0.059} & \textbf{0.006} & \underline{0.557}  & \underline{0.703}  & \underline{0.284}  & \underline{0.360}  \\
    Uni-TSA   & \textbf{0.055}  & \underline{0.007}  & \textbf{0.415} & \textbf{0.525} & \textbf{0.156} & \textbf{0.137} \bigstrut[b]\\
    \hline
    \end{tabular}%
  \label{New England 39-Bus System: State Prediction Capacity Comparison}%
\end{table}%

{\color{red}
\begin{figure}
    \centering
    \subfloat[Stable OC]{\includegraphics[width=0.47\linewidth]{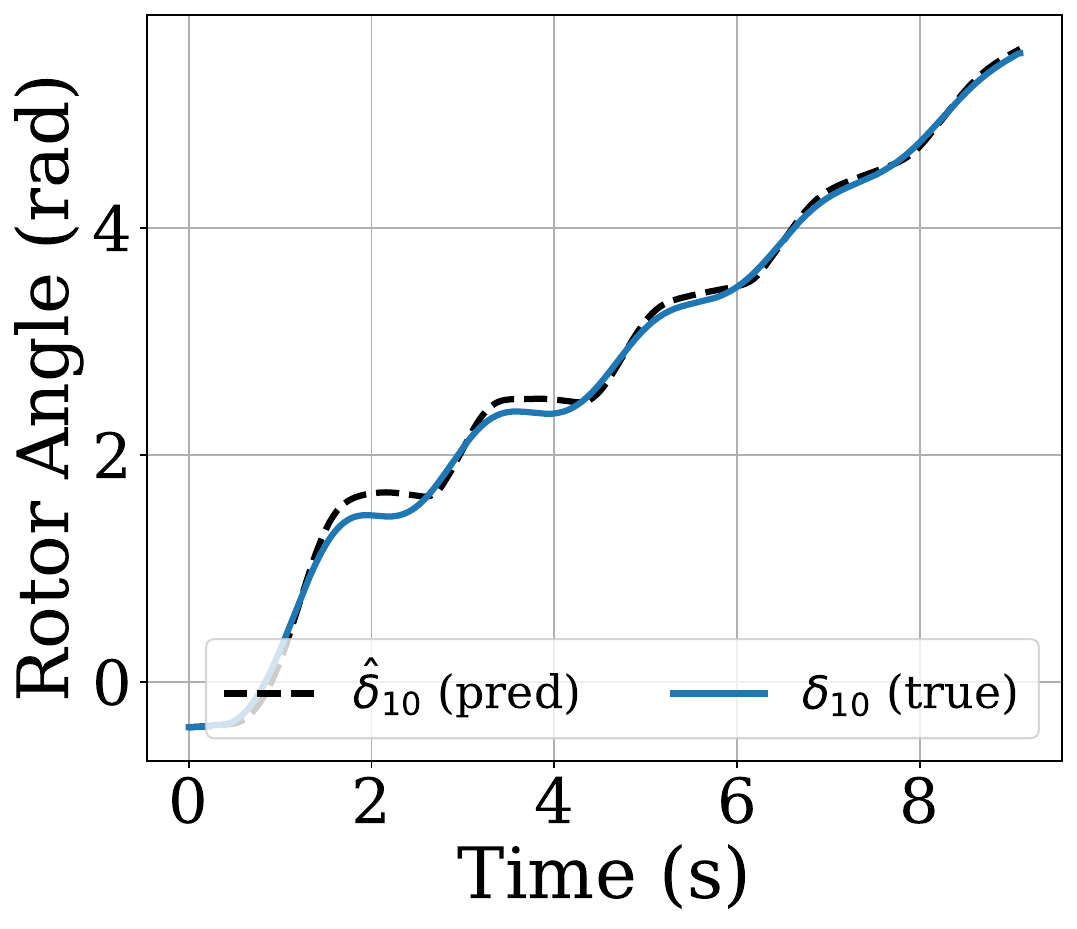}}
    \subfloat[Unstable OC]{\includegraphics[width=0.47\linewidth]{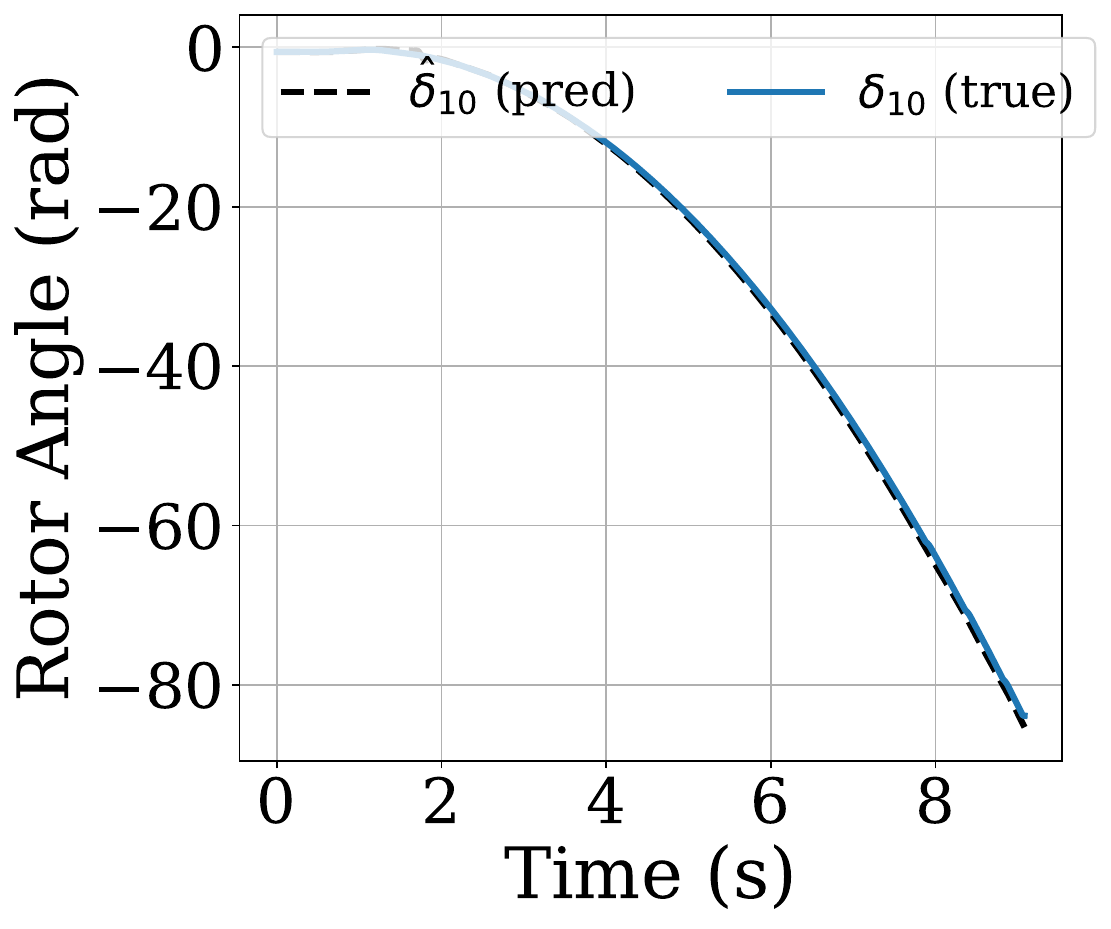}}
    \caption{Representative $\delta$ trajectory comparison for SG 10 (39-bus system). Predictions and ground truths are plotted on the same axes for improved clarity.}
    \label{Representative trajectory comparison for 39-bus SG 10}
\end{figure}
}

\subsection{Zero-shot Generalizability on Unseen Faults}


{\color{black}
This section evaluates the model's zero-shot generalizability to unseen, severe contingencies ($N-2$ and $N-3$) under mixed stability conditions. As detailed in \autoref{New England 39-Bus System: Zero-shot Prediction Capacity Comparison under N-2 Setting} and \autoref{New England 39-Bus System: Zero-shot Prediction Capacity Comparison under N-3 Setting}, all baseline models exhibited a significant performance collapse when confronted with unseen faults. The $MSE_H$ for the RNN-based LSTM and DNR models surged by factors of approximately 13 and 30, respectively, compared to their $N-1$ performance. Even the transformer-based ENC model underperformed, with its $MSE_U$ exceeding 1.5. In contrast, Uni-TSA demonstrated superior zero-shot generalization, with its error on unstable samples ($MAE_U, MSE_U$) remaining remarkably stable and increasing by less than 0.01 relative to the $N-1$ scenario. This robustness provides a substantial advantage, with Uni-TSA reducing $MSE_H$ by nearly {71.05\%--99.17\% for $N-2$ settings} and {84.27\%--99.78\% for $N-3$ settings} relative to the baselines. These performance disparities are rooted in fundamental architectural and training differences. The failure of DNR, for instance, stems not only from the inherent modeling constraints of its RNN architecture but also from its lack of a mechanism to suppress error propagation, leading to severe error accumulation in complex and unstable scenarios. Uni-TSA incorporates a TeaF and SchS scheme specifically to build resilience to its own predictive errors, ensuring stable long-horizon forecasting. Furthermore, when compared to the ENC baseline, Uni-TSA's advantage is twofold. Its larger model scale provides greater expressive capacity, while its causal attention mechanism is better aligned with the nature of online TSA tasks than the ENC's bidirectional attention, explaining its superior performance.
}

\begin{table}[H]
  \centering
  \setlength{\tabcolsep}{3pt} 
  \caption{New England 39-Bus System: Zero-shot Prediction Capacity Comparison under N-2 Setting}
    \begin{tabular}{ccccccc}
    \hline
    Model & $MAE_S$ & $MSE_S$ & $MAE_U$ & $MSE_U$ & $MAE_H$ & $MSE_H$ \bigstrut\\
    \hline
    LSTM  & 0.152  & 0.736  & 2.663  & 21.880  & 1.379  & 19.708  \bigstrut[t]\\
    DNR   & 0.200  & 0.760  & 9.193  & 140.934  & 2.647  & 66.703  \\
    ENC & \underline{0.083}  & \underline{0.010}  & \underline{0.793}  & \underline{1.593}  & \underline{0.781}  & \underline{0.563}  \\
    Uni-TSA   & \textbf{0.042} & \textbf{0.003} & \textbf{0.466} & \textbf{0.498} & \textbf{0.168} & \textbf{0.163} \bigstrut[b]\\
    \hline
    \end{tabular}%
  \label{New England 39-Bus System: Zero-shot Prediction Capacity Comparison under N-2 Setting}%
\end{table}%

\begin{table}[H]
  \centering
    \setlength{\tabcolsep}{3pt} 
  \caption{New England 39-Bus System: Zero-shot Prediction Capacity Comparison under N-3 Setting}
    \begin{tabular}{ccccccc}
    \hline
    Model & $MAE_S$ & $MSE_S$ & $MAE_U$ & $MSE_U$ & $MAE_H$ & $MSE_H$ \bigstrut\\
    \hline
    LSTM  & 0.082  & 0.610  & 2.812  & 27.246  & 1.092  & 15.837  \bigstrut[t]\\
    DNR   & 0.092  & \underline{0.020}  & 5.695  & 142.549  & 2.144  & 70.485  \\
    ENC   & \underline{0.031}  & 0.086  & \underline{0.864}  & \underline{1.567}  & \underline{0.671}  & \underline{0.981}  \\
    Uni-TSA   & \textbf{0.029} & \textbf{0.003} & \textbf{0.457} & \textbf{0.566} & \textbf{0.180} & \textbf{0.154} \bigstrut[b]\\
    \hline
    \end{tabular}%
  \label{New England 39-Bus System: Zero-shot Prediction Capacity Comparison under N-3 Setting}%
\end{table}%

\subsection{Scalability on heterogeneous system}

\begin{figure}
    \centering
    \subfloat[Few-shot scalability]{\includegraphics[width=0.85\linewidth]{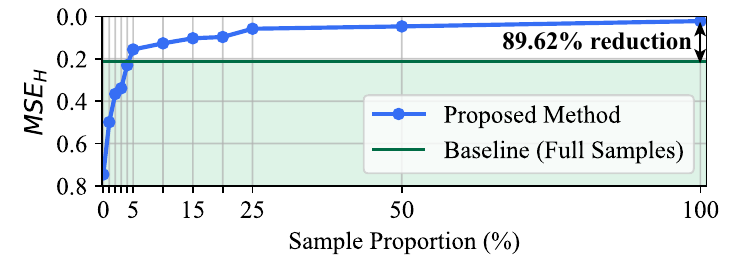}}\\
    \hspace{-1.5em}
\subfloat[Actual $\delta$ in stable OC]{\includegraphics[width=0.46\linewidth]{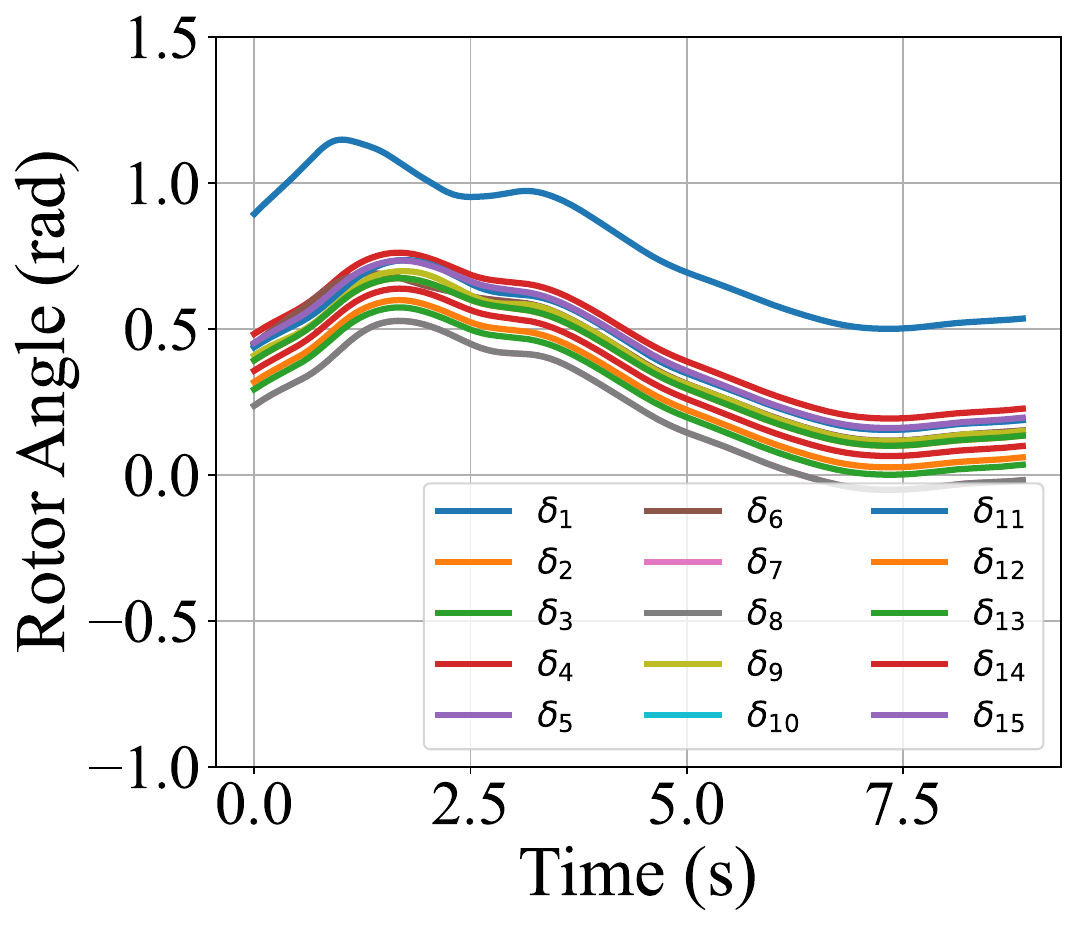}}
\subfloat[Actual $\delta$ in unstable OC]{\includegraphics[width=0.45\linewidth]{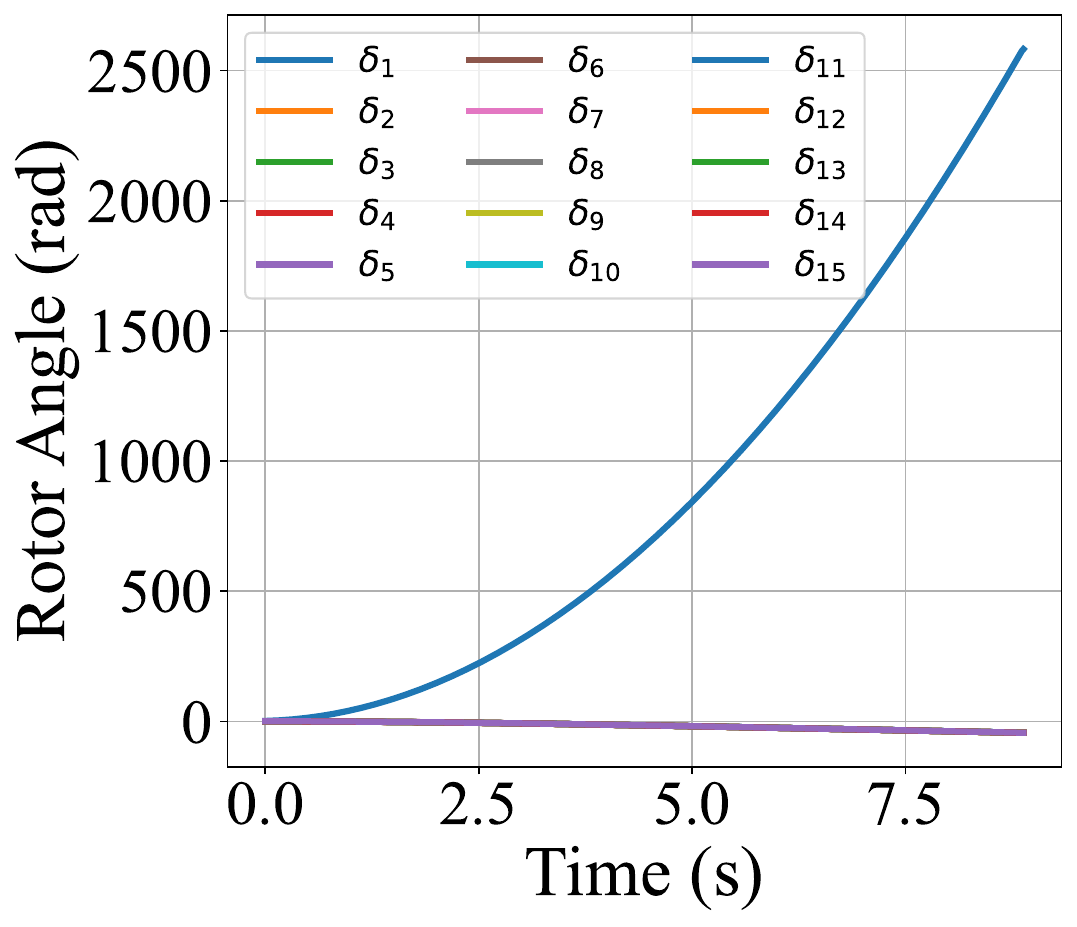}}\\
    \hspace{-1em}
\subfloat[Predicted $\delta$ in stable OC]
{\includegraphics[width=0.46\linewidth]{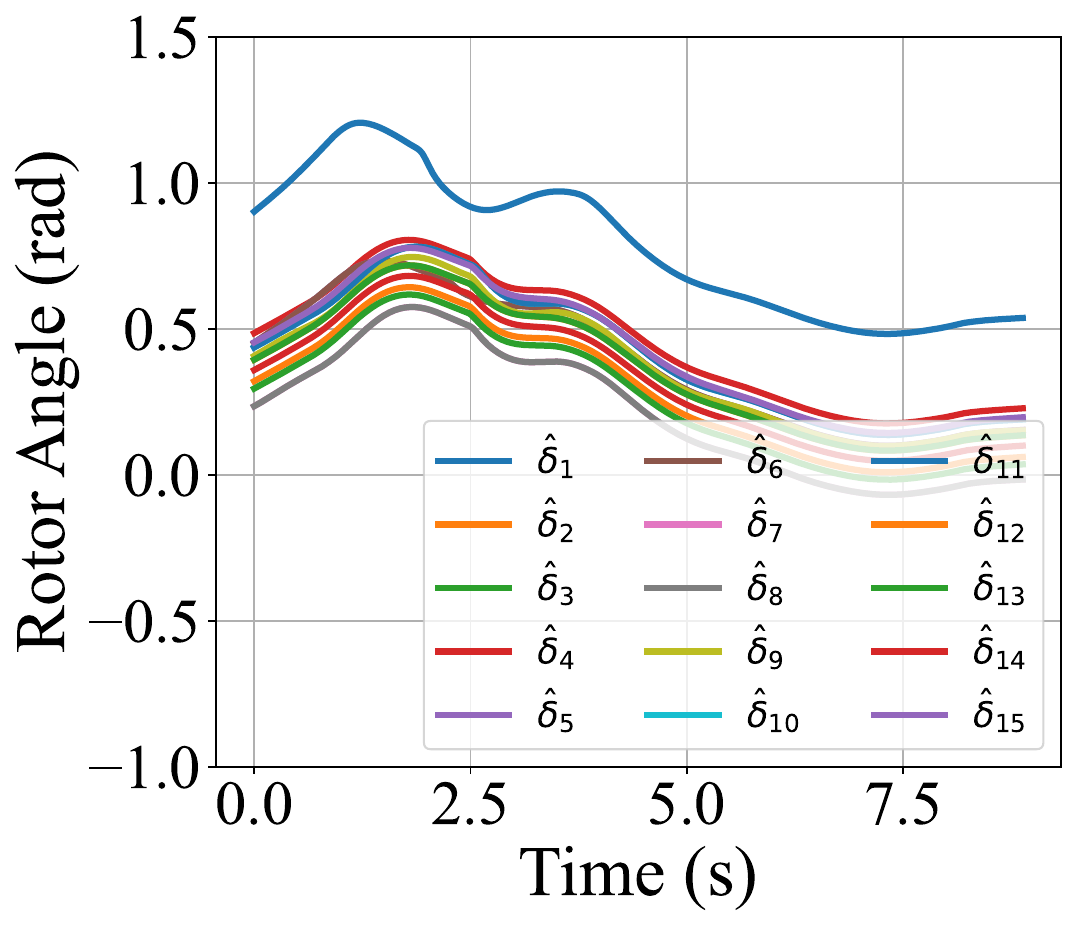}}
\subfloat[Predicted $\delta$ in unstable OC]
{\includegraphics[width=0.45\linewidth]{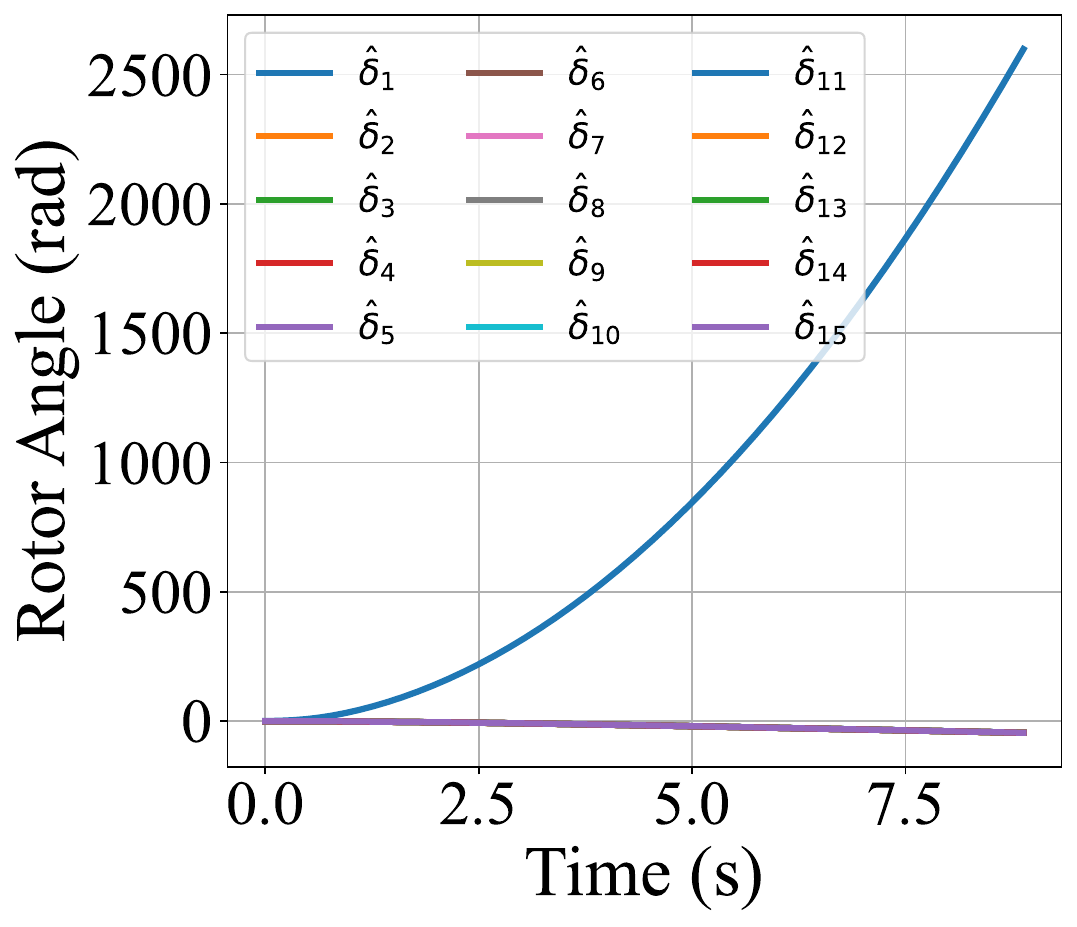}}
    \caption{Iceland 189-bus system: few shot scalability. To facilitate presentation, prediction results for a subset of 15 SGs are displayed.}
    \label{Iceland 189-bus system: few shot scalability.}
\end{figure}

{\color{black}
This section evaluates Uni-TSA's ability to generalize and adapt to a heterogeneous power system with limited target-system data. The model's inherent cross-system potential was initially evaluated through a direct zero-shot test on the $D_{\text{test}}^{189\text{bus}}$ dataset, as its channel independence design permits direct application to systems of varying dimensions. Despite the significant topological and dynamic differences between the systems, Uni-TSA achieved an impressive $MSE_H$ of approximately {0.752}. This zero-shot performance is remarkably strong, even surpassing the in-distribution performance of baseline models like LSTM and DNR on their native New England 39-bus system, as shown in \autoref{New England 39-Bus System: State Prediction Capacity Comparison}.

Furthermore, to assess the model's few-shot learning capability, Uni-TSA pre-trained on the New England 39-bus system was further fine-tuned using varying proportions of $D_{\text{train}}^{189\text{bus}}$, employing the proposed TeaF and SchS schemes. Performance was subsequently evaluated on $D_{\text{test}}^{189\text{bus}}$. As shown in \autoref{Iceland 189-bus system: few shot scalability.}, the $MSE_H$ of Uni-TSA rapidly decreases with increasing proportions of the Iceland 189-bus system fine-tuning data, and trends for other metrics are generally consistent. For a stringent comparison, the ENC model was trained on the {full} $D_{\text{train}}^{189\text{bus}}$ dataset to serve as an "expert" baseline for the target system. The results demonstrate that Uni-TSA achieves performance comparable to this fully-trained expert after fine-tuning on only 4\%--5\% of the available target-system data. This remarkable data efficiency is visually confirmed by the predicted trajectories from the Uni-TSA fine-tuned by 5\% samples from $D_{\text{train}}^{189\text{bus}}$ in \autoref{Iceland 189-bus system: few shot scalability.} {\color{red}and in \autoref{Representative trajectory comparison for 189-bus SG 4}, where predictions and ground truths for SG 2 are plotted on the same axes for improved clarity.} Furthermore, when fine-tuned on the complete Iceland 189-bus system dataset, Uni-TSA reduces the $MSE_H$ by nearly {89.62\%} compared to the expert ENC model, underscoring its superior architectural capacity. This cross-system performance introduces the potential to replace the costly, expert-intensive process of building custom-built models for large-scale systems {by enabling} an efficient methodology where a foundational model is trained on smaller systems and subsequently adapted with minimal target-system data. Beyond this immediate economic advantage, our findings reveal a clear path for true performance scalability. The model is designed to continually enhance its multi-system, multi-scenario proficiency by incorporating data from an ever-expanding corpus of power systems, paving the way for a single, continually evolving foundation TSA model.
}

{\color{red}
\begin{figure}
    \centering
    \subfloat[Stable OC]{\includegraphics[width=0.47\linewidth]{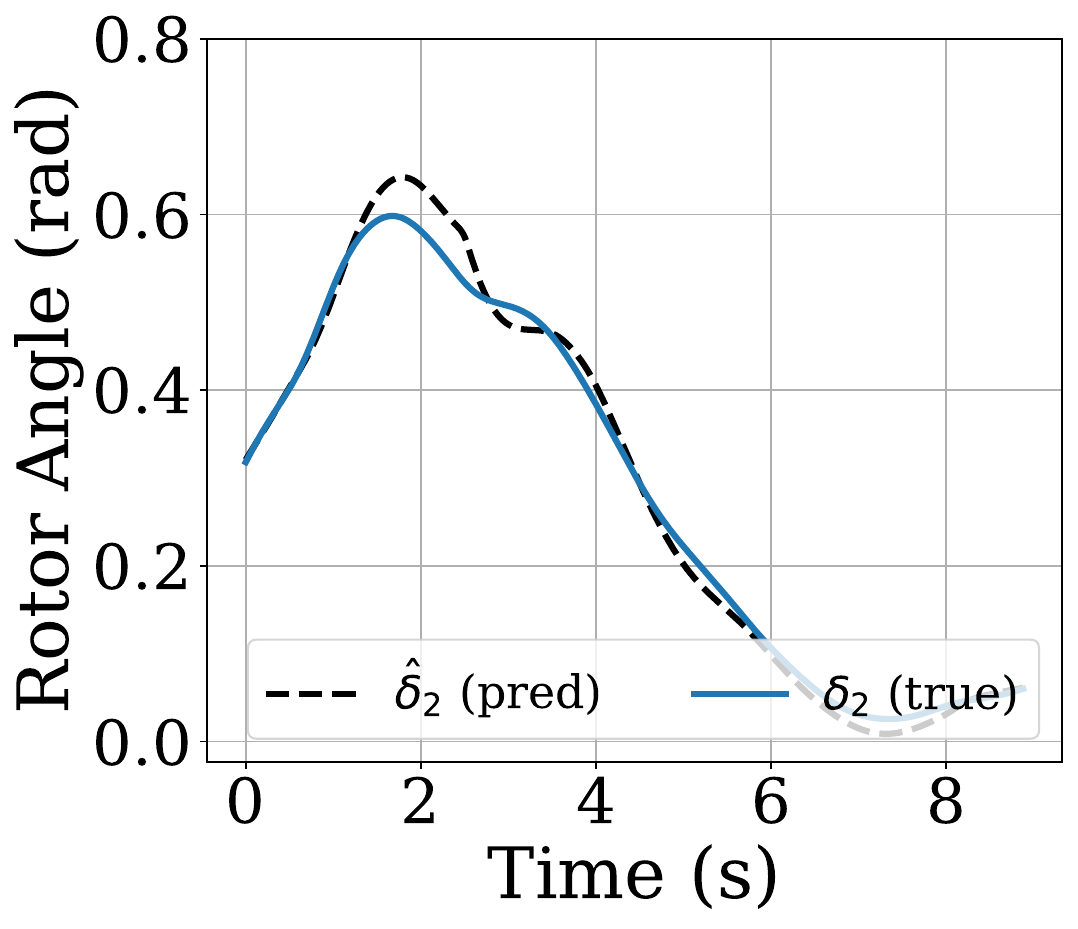}}
    \subfloat[Unstable OC]{\includegraphics[width=0.47\linewidth]{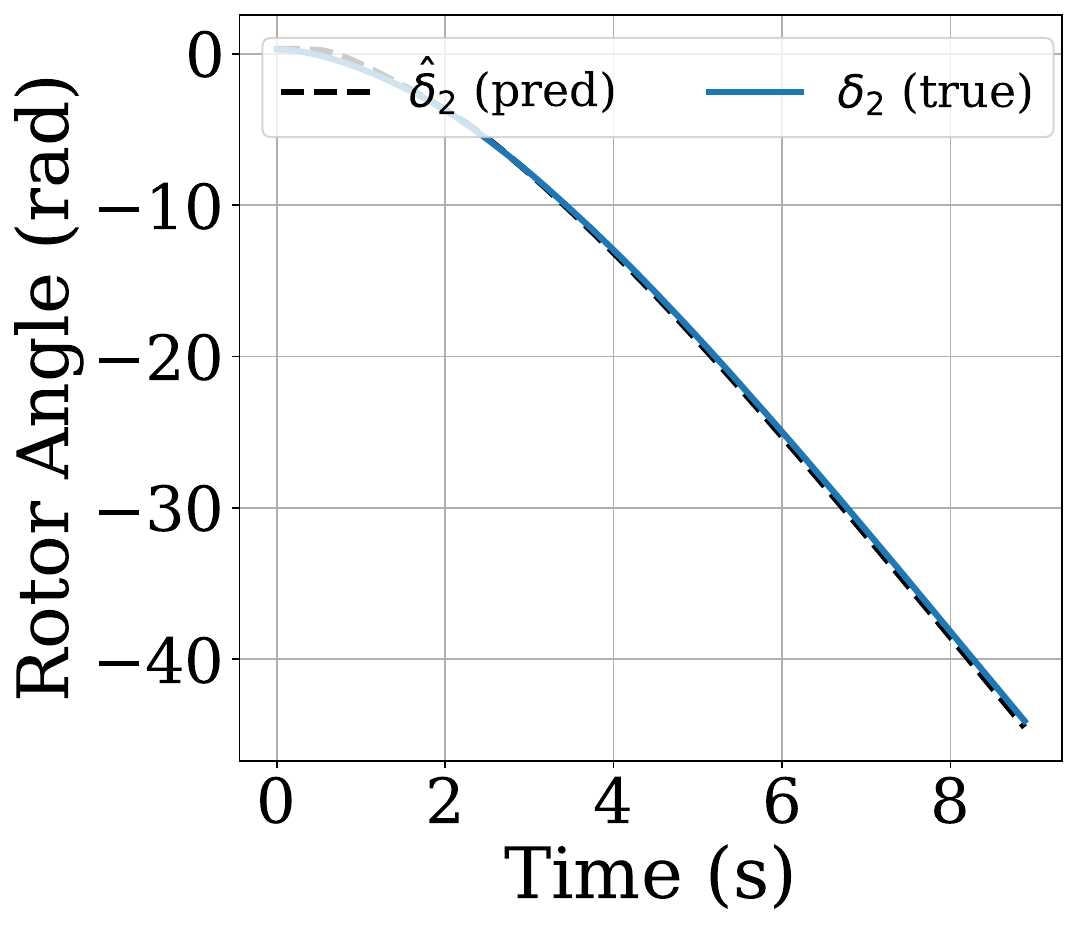}}
    \caption{Representative $\delta$ trajectory comparison for SG 2 (189-bus system). Predictions and ground truths are plotted on the same axes for improved clarity.}
    \label{Representative trajectory comparison for 189-bus SG 4}
\end{figure}
}





\subsection{Ablation Study} \label{Ablation Study}
{\color{black}
To assess the individual contributions of the dual-phase training strategy (TeaF and SchS) and the temporal patch, a series of ablation studies were performed, with results summarized in \autoref{Ablation study comparing the performance of full proposed model against variants with removed components.}. Model variants, denoted ''w/o'' for the exclusion of a component, were trained on the $D_{\text{train}}$ and $D_{\text{val}}$. Their performance was subsequently evaluated across all test sets ($D_{\text{test}}^{N-1}$, $D_{\text{test}}^{N-2}$, $D_{\text{test}}^{N-3}$, and $D_{\text{test}}^{189\text{bus}}$), with a 5\% fine-tuning step on $D_{\text{train}}^{189\text{bus}}$ applied specifically for the final cross-system test. The results confirm that all components are integral to the model's performance. The removal of TeaF and SchS, which form the dual-phase training strategy, led to the most significant performance degradation, with average error increases of nearly 68.76\% and 41.16\%, respectively, highlighting their importance in robustly modeling dynamics and suppressing cumulative errors. Lastly, removing the patch operation resulted in a nearly 21.26\% performance decline, underscoring its importance in helping the attention mechanism model rich and informative local dynamic patterns.}

\begin{table}[htbp]
  \centering
  \caption{Ablation Study Comparing the Performance of Full Proposed Model against Variants with Removed Components.}
  
  \begin{tabular}{c|cc|cc}
    \hline
    \multirow{2}[2]{*}{Model} & \multicolumn{2}{c|}{$D_{\text{test}}^{N-1}$} & \multicolumn{2}{c}{$D_{\text{test}}^{N-2}$} \bigstrut[t]\\
    \cline{2-5}
          & $MAE_H$ & $MSE_H$ & $MAE_H$ & $MSE_H$ \bigstrut[b]\\
    \hline
    w/o Teacher Forcing & 0.194  & 0.296  & 0.241  & 0.329  \bigstrut[t]\\
    w/o Schedule Sampling & 0.189  & 0.178  & 0.192  & 0.258  \\
    w/o Patch & \underline{0.161}  & \underline{0.145}  & \underline{0.192}  & \underline{0.230}  \\
    Uni-TSA & \textbf{0.156} & \textbf{0.137} & \textbf{0.168} & \textbf{0.163} \bigstrut[b]\\
    \hline
  \end{tabular}

  \begin{tabular}{c|cc|cc}
    \hline
    \multirow{2}[2]{*}{Model} & \multicolumn{2}{c|}{$D_{\text{test}}^{N-3}$} & \multicolumn{2}{c}{$D_{\text{test}}^{189bus}(5\%)$} \bigstrut[t]\\
    \cline{2-5}
          & $MAE_H$ & $MSE_H$ & $MAE_H$ & $MSE_H$ \bigstrut[b]\\
    \hline
    w/o Teacher Forcing & 0.310  & 0.222  & 0.113  & 0.299  \bigstrut[t]\\
    w/o Schedule Sampling & 0.267  & 0.202  & 0.138  & 0.210  \\
    w/o Patch & \underline{0.198}  & \underline{0.173}  & \underline{0.114}  & \underline{0.196}  \\
    Uni-TSA & \textbf{0.180} & \textbf{0.154} & \textbf{0.072} & \textbf{0.156} \bigstrut[b]\\
    \hline
  \end{tabular}
  
\label{Ablation study comparing the performance of full proposed model against variants with removed components.}
\end{table}

\subsection{Validation on Angular Speed}\label{angular speed test}
{\color{black}
Recognizing the operational need for comprehensive state variable monitoring, this subsection evaluates the proposed model's predictive performance for rotor angular speed, with all predictions presented in per-unit (p.u.) values. As summarized in \autoref{Angular Speed Prediction Performance Comparison}, the $MSE_H$ results indicate that the proposed model demonstrates an even more substantial performance advantage in rotor speed prediction compared to its rotor angle prediction capabilities. Across the four primary evaluation tasks, the proposed model achieves performance improvements of up to {two orders of magnitude} over baselines. For instance, when trained with the complete $D_{\text{train}}^{\text{189bus}}$, the proposed approach reduced the $MSE_H$ by nearly {95.95\%} relative to the leading baseline model (ENC) on $D_{\text{test}}^{\text{189bus}}$. Furthermore, in a few-shot learning scenario on this system, the proposed method matched the performance of the fully trained ENC model while utilizing only approximately 3\% of the training data---a smaller proportion than required for the rotor angle prediction task. These findings underscore the proposed method's efficacy and versatility in predicting various critical state variables.}

To further explore alternative data processing strategies, an additional experiment,
Uni-TSA (Unit-wise), is conducted, as shown in the last row of \autoref{Angular Speed Prediction Performance Comparison}.
While the original Uni-TSA processes state variables independently to maximize
cross-system adaptability, we investigate an alternative approach inspired by \cite{tan2025bayesian},
where each synchronous generator (SG) is treated as an integrated unit.
In this unit-wise formulation, the state trajectories of each SG (e.g., rotor angle
$\delta$ and speed $\omega$) are concatenated and jointly represented as multivariate
time series, enabling the model to capture intra-unit correlations while maintaining
inter-system generality.
As shown in \autoref{Angular Speed Prediction Performance Comparison}, the unit-wise variant achieves a $95.33\%$ reduction in
$\mathrm{MSE}_H$ on the full $D_{\text{test}}^{189\text{bus}}$ and matches the fully
trained ENC baseline using only $3\%$ of the training data.
These results demonstrate that unit-wise temporal modeling provides a viable
alternative for capturing variable dependencies while preserving cross-system
flexibility.

\begin{table}[htbp]
    \centering
      \renewcommand{\arraystretch}{1.1}
    \caption{Angular Speed Prediction Performance Comparison by $MSE_H$}
      \begin{tabular}{ccccc}
      \hline
      Model & $D_{\text{test}}^{N-1}$ & $D_{\text{test}}^{N-2}$ & $D_{\text{test}}^{N-3}$ & $D_{\text{test}}^{\text{189bus}}(100\%)$ \\ [1ex]
      \hline
      LSTM  & 0.0827  & 0.0904  & 0.1002  & 0.2036 \\
      DNR   & {0.0395}  & {0.0457}  & {0.0389}  & 0.1235 \\
      ENC   & 0.0427  & 0.0478  & 0.0529  & {0.1136} \\
      Uni-TSA & {0.0012} & {0.0008} & {0.0030} & {0.0056 (3\%)} \\    
      \hdashline
      \noalign{\vspace{2pt}}
      \makecell{Uni-TSA \\ (Unit-wise)} & {0.0010} & {0.00007} & {0.0034} & {0.0053 (3\%)}\\
      \hline
      \end{tabular}%
    \label{Angular Speed Prediction Performance Comparison}%
  \end{table}%


\subsection{Training and Inference Cost Comparison}

{\color{black}
This section analyzes the computational efficiency of the proposed Uni-TSA, comparing it against the ENC baseline and the traditional TDS on the large-scale 189-bus system. We assess efficiency using two key metrics:  1) the training cost per step, defined as a single forward and backward propagation pass on one batch, and 2) the inference time to predict the complete trajectories for a single operating condition (OC). The detailed results of this comparison are presented in \autoref{Training Parameters and Training/Inference Cost Comparison}.
\subsubsection{Training Efficiency}
The remarkable training efficiency of the large-scale Uni-TSA, which remains comparable to much smaller models, is almost entirely attributable to its {parameter-efficient fine-tuning} approach. The strategy of freezing the main T-blocks and training only 0.69\% of the total parameters significantly reduces the computational complexity associated with the most intensive phase of training, namely the backward pass and parameter update. This allows the model to be fine-tuned with exceptional speed, evidenced by a per-step training time on the large 189-bus system that is only negligibly slower (0.004s) than the compact ENC baseline.
\subsubsection{Inference Speed} In the inference stage, Uni-TSA exhibits a significant speed advantage, proving faster than even the smaller ENC baseline on the 189-bus system. This high efficiency stems from three core design principles. First, the {patch and stride operations} strategically reduce the input sequence length for the attention mechanism, which is instrumental in mitigating the quadratic computational complexity ($\mathcal{O}(N^2)$) inherent in long-sequence modeling. Second, the {causal attention} mechanism results in a sparse attention weight matrix where most elements are zero, thereby simplifying the computational process of aggregating patch representations. Most critically, the {online application design} reframes the problem by transforming all $n_x$ system channels into the batch dimension. This enables the model to process the entire dynamic trajectory for a given operating condition simultaneously while fully leveraging the massive parallel processing capabilities of GPUs on large batches. Consequently, Uni-TSA not only outperforms the ENC model but also achieves an inference speedup of approximately 96.88\% over traditional TDS. These results underscore the framework's strong potential for efficient, real-time trajectory prediction.}

{\color{red}
It is worth clarifying that the reported 0.57~s corresponds to the end-to-end inference time under our specific hardware and implementation settings for predicting the complete post-fault trajectory over a 9~s horizon for one OC, and it is not a fundamental lower bound. In practical deployment, standard GPU inference optimizations (e.g., mixed precision/quantization and optimized runtime backends) can further reduce latency. Moreover, for many unstable cases, system instability can often be identified from the early-stage response without requiring full-horizon prediction, implying that the effective response time for triggering emergency control can be substantially shorter than 0.57~s when a shorter screening horizon is adopted. Finally, our online application design batches all channels for concurrent prediction via GPU parallelization, so the inference time does not scale linearly with system size in the same way as time-domain simulation; as system scale increases and TDS becomes slower, the relative speed advantage of Uni-TSA is expected to be even more pronounced.}

\begin{table}
  \centering
  \caption{Training Parameters and Training/Inference Cost Comparison in the Modified Iceland 189-Bus System}
  \setlength{\tabcolsep}{3pt}
    \begin{tabular}{ccccc}
    \hline
    Method & \makecell{Trainable\\Params} & \makecell{Trainable Params\\Percentages} & \makecell{Training\\(s/step)} & \makecell{Inference\\(s/OC)} \bigstrut[t]\bigstrut[b]\\
    \hline
    TDS   & -     & -     & -     & 18.25  \bigstrut[t]\\
    ENC   & 9.6M  & 100\%   & 0.034  & 0.59  \\
    Uni-TSA   & 858K  & 0.69\%  & 0.038  & 0.57  \bigstrut[b]\\
    \hline
    \end{tabular}%
  \label{Training Parameters and Training/Inference Cost Comparison}%
\end{table}%

\subsection{Generic Validation of Pre-trained Parameters}\label{Qualitative Analysis of Pretrained Parameters in LLM}
{\color{blue}
\subsubsection{Performance Gain Attributable to Pre-training}
Pre-trained T-blocks are a key factor behind the cross-scenario generalization of Uni-TSA.
To quantify their contribution, we further compare three settings on \(D_{\text{test}}^{N-1}\) under the same GPT-2 backbone: a randomly initialized model with freeze-and-finetune, a pure Transformer model trained from scratch, and the proposed Uni-TSA with pretrained parameters and freeze-and-finetune.
As shown in \autoref{Comparison of Model Performance and Token Similarity with and without Pre-trained Parameters}, the pure Transformer setting still exhibits much larger prediction errors than Uni-TSA. Specifically, Uni-TSA reduces \(MAE_H\) from 1.023 to 0.156 and \(MSE_H\) from 18.091 to 0.137. These results indicate that the performance gain cannot be attributed to the Transformer architecture alone, and that the pretrained T-blocks within pre-trained generative Transformers provide transferable sequence modeling priors for capturing transient power-system dynamics.

\subsubsection{Explaining Performance Gains}

After establishing the performance gap, we further investigated whether pre-trained T-blocks preserve the representational properties described in \ref{Generic Feature Extraction of Transformer Block}.  
To align this analysis with the subsequent mechanism discussion, we used two indicators: \emph{feature stability} and \emph{feature co-directionality} (see \autoref{Comparison of Model Performance and Token Similarity with and without Pre-trained Parameters}). 

\emph{Feature stability} characterizes cross-layer consistency, i.e., whether the same patch remains similarly encoded as depth increases. We measure it by the mean cosine similarity between the representations of each patch in the last two Transformer layers. Uni-TSA attains 0.893, compared with 0.722 for the pure Transformer counterpart, indicating more stable feature propagation across layers.  

\emph{Feature co-directionality} characterizes intra-sequence alignment at a fixed layer. For each sequence, we compute pairwise cosine similarities among patch representations in the final layer and report the proportion of patch pairs with cosine similarity larger than 0.8. Uni-TSA reaches 71.44\%, versus 48.47\% for the pure Transformer counterpart.  

Overall, these results support the mechanism analysis in \ref{Generic Feature Extraction of Transformer Block} and indicate that pre-trained T-blocks preserve transferable sequence representation properties for transient dynamics modeling. 

From a physical perspective, transient trajectories in power systems with synchronous generators are governed by the same basic physics mechanisms, e.g., the swing equation coupled with network algebraic constraints. Because these underlying physical laws are universal, a model only needs to extract stable and well-aligned temporal representations to effectively capture these shared dynamical patterns across different scenarios and systems. 

However, developing such generic and robust feature extraction capabilities from scratch is notoriously data-hungry. 
Since the supervised TSA datasets are far smaller in scale and diversity than the massive corpora used in language pre-training, a pure Transformer trained from scratch struggles to autonomously form equally stable sequence representations (as evidenced by its feature stability dropping to 0.722 and co-directionality to 48.47\%). In contrast, the pre-trained T-blocks leverage their prior sequence modeling capabilities to robustly extract these consistent temporal features (attaining a stability of 0.893 and a co-directionality of 71.44\%), which directly explains their superior performance in cross-scenario and cross-system generalization.

}

\begin{table}
  \renewcommand{\arraystretch}{1.1} 
    \centering
    \caption{Comparison of Model Performance with and without Pre-trained Parameters}
    \begin{tabular}{ccccc}
    \hline
        Backbone & \makecell{Pre-trained\\parameters} & \makecell{Freeze-and-\\finetune} & $MAE_H$ & $MSE_H$ \\ \hline
        GPT-2 & $\times$ & $\checkmark$ & 1.134  & 23.220  \\
        GPT-2 & $\times$ & $\times$ & 1.023 & 18.091 \\
        GPT-2 & $\checkmark$ & $\checkmark$ & \textbf{0.156}  & \textbf{0.137} \\ \hline
        \multirow{2}{*}{Backbone} & \multirow{2}{*}{\makecell{Pre-trained\\parameters}} & \multirow{2}{*}{\makecell{Freeze-and-\\finetune}} & Feature & Feature \\ 
         &  &  & stability & co-direction ($>$0.8) \\ \hline
        GPT-2 & $\times$ & $\checkmark$ & 0.673 & 46.78\% \\
        GPT-2 & $\times$ & $\times$ & 0.722 & 48.47\% \\
        GPT-2 & $\checkmark$ & $\checkmark$ & \textbf{0.893} & \textbf{71.44\%} \\ \hline
    \end{tabular}
    \label{Comparison of Model Performance and Token Similarity with and without Pre-trained Parameters}
\end{table}

{\color{red}
\subsection{Backbone Ablation with Qwen3-0.6B}
To further validate that the proposed framework is backbone agnostic, we instantiated Uni-TSA with an alternative pretrained decoder-only Transformer, Qwen3-0.6B. While both the original GPT backbone and Qwen3-0.6B adopt a decoder-only architecture, Qwen3-0.6B additionally uses a Mixture-of-Experts design. For a controlled comparison, we kept the data processing pipeline unchanged, including channel-independence, sample-wise normalization, and patching. We also used the same training protocol, including TeaF and SchS, and the same evaluation setup. In other words, we changed the backbone while holding all other components fixed.

Table~\ref{Estimated MSE_H Comparison with Qwen3-0.6B MoE Backbone} reports the resulting $MSE_H$ scores. The Qwen3-0.6B (MoE) variant yields consistently lower errors than the original Uni-TSA across all four benchmarks, including $D_{\text{test}}^{N-1}$, $D_{\text{test}}^{N-2}$, $D_{\text{test}}^{N-3}$, and $D_{\text{test}}^{\text{189bus}}(5\%)$.

These results lead to two observations. First, the Uni-TSA framework transfers naturally to other pretrained large-model backbones beyond GPT within the decoder-only family, which supports its architectural compatibility and portability. Second, increasing the effective model capacity via an MoE backbone appears to strengthen sequence representations and nonlinear dynamics modeling, improving long-horizon trajectory prediction in both unseen-fault and cross-system settings. Overall, scaling the backbone provides a practical route to further enhance TSA universality and prediction accuracy.
\begin{table}[htbp]
  \centering
  \caption{Estimated $MSE_H$ comparison after backbone replacement with Qwen3-0.6B}
  \renewcommand{\arraystretch}{1.1}
  \begin{tabular}{ccccc}
    \hline
    Model & $D_{\text{test}}^{N-1}$ & $D_{\text{test}}^{N-2}$ & $D_{\text{test}}^{N-3}$ & $D_{\text{test}}^{\text{189bus}}(5\%)$ \\
    \hline
    Uni-TSA (GPT) & 0.137 & 0.163 & 0.154 & 0.156 \\
    Uni-TSA (Qwen3-0.6B MoE) & \textbf{0.126} & \textbf{0.149} & \textbf{0.141} & \textbf{0.142} \\
    \hline
  \end{tabular}
  \label{Estimated MSE_H Comparison with Qwen3-0.6B MoE Backbone}
\end{table}
}

{\color{red}
\subsection{Cross-System Scaling on the IEEE 68-bus System and the IEEE 118-bus System}\label{sec:scaling_68_118}
To further strengthen the empirical justification of \textit{universality} within the IEEE benchmark regime and to better illustrate the scaling trend beyond the IEEE-39/189 setting, we additionally evaluate Uni-TSA on two additional benchmark systems, namely the IEEE 68-bus system and the IEEE 118-bus system, under $N\!-\!1$ contingencies. The transient trajectories are generated following the same simulation pipeline described in Sec.~4.1 (\textit{Data Generation and Model Settings}). This experiment is intended to provide additional benchmark evidence on cross-system scalability and data-efficiency, rather than to claim that IEEE-39 data alone can guarantee direct zero-shot generalization to real utility-scale grids with substantially higher heterogeneity and renewable penetration.
Specifically, we generate 14,400 trajectories for the IEEE 68-bus system (16 SGs) and 17,100 trajectories for the IEEE 118-bus system (19 SGs), corresponding to 900 operating conditions (OCs) in total, and split each dataset into $D_{\text{train}}^{\text{target}}$ and $D_{\text{test}}^{\text{target}}$ with a 2:1 ratio (9,600/4,800 for the IEEE 68-bus system and 11,400/5,700 for the IEEE 118-bus system). In particular, the test set comprises 300 OCs, and the OCs are generated with a stable/unstable proportion of 7:3.

In addition to trajectory-prediction errors, we further report stability-relevant classification metrics derived from the predicted trajectories. Specifically, we define a transient stability index (TSI) based on the maximum rotor-angle separation with respect to the center-of-inertia (COI) \cite{shen2025physics,shen2025physics_argumented}. For an OC $k$, let $\delta_{i,k}(t)$ denote the rotor angle of generator $i$, and define
\begin{align}
\Delta\delta_k(t) &= \max_{i}\left|\delta_{i,k}(t)-\delta_{\text{COI},k}(t)\right|,\\
\mathrm{TSI}_k &= \max_{t\in [t_0,\,T]}\Delta\delta_k(t),
\end{align}
where $[t_0,T]$ is the post-fault simulation horizon. Given a predefined stability threshold $\delta_{\text{th}}$ (set to $\pi$), the ground-truth stability label is
\begin{align}
y_k=\mathbb{I}\left(\mathrm{TSI}_k>\delta_{\text{th}}\right)\in\{0,1\},
\end{align}
where $y_k=1$ indicates an unstable (loss-of-synchronism) OC and $y_k=0$ indicates a stable OC.
The predicted label $\hat{y}_k$ is obtained by applying the same rule to the predicted trajectories. We then report the classification accuracy (Acc) and the recall of the unstable class (Recall-U) as
\begin{align}
\mathrm{Acc} &= \frac{1}{N}\sum_{k=1}^{N}\mathbb{I}\left(\hat{y}_k=y_k\right),\\
\mathrm{Recall\text{-}U} &= \frac{\sum_{k=1}^{N}\mathbb{I}\left(\hat{y}_k=1\right)\mathbb{I}\left(y_k=1\right)}{\sum_{k=1}^{N}\mathbb{I}\left(y_k=1\right)},
\end{align}
where $N$ is the number of OCs in the test set.

We consider three transfer settings: (i) \textit{zero-shot} evaluation, where the model is fine-tuned only on the IEEE 39-bus system and directly tested on the target system; (ii) \textit{few-shot} adaptation on the target system with limited data ratios (10\% and 20\%); and (iii) \textit{full-data} fine-tuning on the target system. For comparison, we also report the performance of the strongest baseline, ENC, trained with the full target-system data as an expert model.

Table~\ref{tab:scaling_68_118} summarizes the results. On the IEEE 68-bus system, the zero-shot transfer from the IEEE 39-bus system attains an $MSE_H$ of 0.620, and few-shot fine-tuning with only 10\% target data already outperforms the full-data ENC expert (0.122 vs. 0.135). With 20\% target data, the error is further reduced to 0.106, and full-data fine-tuning reaches 0.082. A similar trend is observed on the IEEE 118-bus system, where $MSE_H$ decreases from 0.710 (zero-shot) to 0.136 with 10\% target data, surpassing the ENC expert trained on the full target data (0.149), and further improves to 0.113 (20\%) and 0.089 (full-data). Notably, the conventional RNN baselines (LSTM and DNR) exhibit orders-of-magnitude larger trajectory errors and markedly lower unstable-recall, indicating their limited robustness in mixed stable/unstable post-fault settings.

For a supplementary comparison on the shared task of transient stability classification, we further include the Potential Energy Boundary Surface (PEBS) method as a representative classical direct-method baseline \cite{chiang1988pebs}. PEBS is an energy-function-based method that approximates the post-fault stability boundary by locating the first local maximum of the potential energy along the fault-on trajectory, thereby avoiding explicit UEP computation and retaining attractive screening efficiency for large contingency sets. Since PEBS outputs only stability decisions rather than predicted trajectories, $MSE_H$ is not applicable. As shown in Table~\ref{tab:scaling_68_118}, PEBS attains Acc/Recall-U of 75.0\%/76.7\% on the IEEE 68-bus system and 77.7\%/67.8\% on the IEEE 118-bus system. Its unstable-class recall remains higher than those of the conventional RNN baselines, showing that PEBS can still capture a meaningful portion of critical unstable cases. However, its relatively modest overall accuracy suggests a conservative decision tendency, i.e., improved unstable-case coverage is achieved at the cost of more false alarms on stable OCs. By contrast, Uni-TSA consistently delivers both higher Acc and higher Recall-U, indicating a more balanced and reliable stability discrimination capability on these benchmark systems, while simultaneously providing full trajectory-level prediction and cross-system transfer capability. 

Overall, these benchmark results indicate that Uni-TSA can be efficiently adapted to other benchmark systems with limited target-system data, while maintaining a clear improvement trend as the fine-tuning data increases. Extending the same framework to real utility-scale systems will require expanding the training corpus to cover larger-scale and renewable-rich dynamics, which we identify as future work.

\begin{table}[H]
  \centering
  \caption{Cross-system scaling evaluation on the IEEE 68-bus system and the IEEE 118-bus system under $N\!-\!1$ contingencies.}
  \renewcommand{\arraystretch}{1.1}
  \setlength{\tabcolsep}{3pt}
  \small
  \resizebox{\linewidth}{!}{%
  \begin{tabular}{llccc}
    \hline
    Target system & Method/Setting & $MSE_H$ & Acc (\%) & Recall-U (\%) \\
    \hline
    \multirow{8}{*}{IEEE 68-bus}
    & Uni-TSA (Zero-shot) & 0.620 & 80.0 & 70.0 \\
    & Uni-TSA (Few-shot, 10\%) & 0.122 & 95.0 & 95.6 \\
    & Uni-TSA (Few-shot, 20\%) & 0.106 & 96.0 & 96.7 \\
    & Uni-TSA (Full-data) & \textbf{0.082} & \textbf{97.0} & \textbf{97.8} \\
    & ENC (Full-data) & 0.135 & 94.7 & 91.1 \\
    & LSTM (Full-data) & 12.876 & 89.0 & 55.6 \\
    & DNR (Full-data) & 9.516 & 90.3 & 60.0 \\
    & PEBS & - & 75.0 & 76.7 \\
    \hline
    \multirow{8}{*}{IEEE 118-bus}
    & Uni-TSA (Zero-shot) & 0.710 & 78.0 & 68.9 \\
    & Uni-TSA (Few-shot, 10\%) & 0.136 & 94.3 & 94.4 \\
    & Uni-TSA (Few-shot, 20\%) & 0.113 & 95.3 & 96.7 \\
    & Uni-TSA (Full-data) & \textbf{0.089} & \textbf{96.3} & \textbf{97.8} \\
    & ENC (Full-data) & 0.149 & 94.0 & 90.0 \\
    & LSTM (Full-data) & 15.083 & 88.0 & 51.1 \\
    & DNR (Full-data) & 11.809 & 89.3 & 55.6 \\
    & PEBS & - & 77.7 & 67.8 \\
    \hline
  \end{tabular}%
  }
  \label{tab:scaling_68_118}
\end{table}
}

\section{Conclusion}

{\color{black}
This paper introduces Uni-TSA, a universal framework that leverages a pre-trained generative Transformer to achieve robust generalization across diverse OCs, unseen faults, and heterogeneous systems with a single architecture. This framework comprises a novel data processing pipeline featuring channel-independence, sample-wise normalization, and temporal patching to handle mixed-stability and long sequences from heterogeneous systems. Moreover, a parameter-efficient, freeze-and-finetune strategy for the augmented GPT architecture is designed to preserve generic sequence feature extraction from pre-trained T-blocks while allowing for TSA task adaptation. Furthermore, a two-stage fine-tuning scheme that combines TeaF with SchS is devised to mitigate cumulative prediction errors and ensuring long-horizon reliability. Case studies on two benchmark systems demonstrate that the proposed approach exhibits exceptional universality, low training and inference costs, and a considerable degree of interpretability, establishing it as a promising foundation for real-time TSA.}

{\color{red}\subsection{Scope and Limitations}}
{\color{red}The proposed framework is validated on several standard IEEE test systems with different scales and operating characteristics. Regarding \textit{system size}, Uni-TSA has demonstrated transferability across multiple benchmark systems. Extending it to utility-scale grids will mainly depend on systematic simulations on larger and more heterogeneous systems to construct representative training and evaluation datasets. Regarding the \textit{diversity of operating conditions}, the present experiments include randomized load variations and wind-induced stochasticity. Broader coverage will require additional renewable scenarios and more general non-optimal pre-fault states generated beyond OPF-centered settings. Regarding \textit{disturbance types}, the current validation covers variation in fault location and contingency severity, but mainly under three-phase short-circuit faults. Extending the framework to a broader disturbance space will require incorporating additional symmetrical and asymmetrical faults, such as single line-to-ground and line-to-line faults. Regarding \textit{model assumptions}, Uni-TSA follows a measurement-driven setting in which topology and parameter effects are learned implicitly from post-fault trajectories instead of being encoded explicitly. Relaxing this assumption will require introducing more explicit structural priors into the modeling framework.}

{\color{red}\subsection{Future Work}
As future work, embedding specialized power system mechanism knowledge into the model 
structure or training objective (e.g., by adding physics informed regularization such as DAE/swing consistency 
residual losses) remains a promising direction worthy of further exploration, which may further enhance the 
framework's interpretability and generalization under data-scarce or out-of-distribution scenarios. In addition, 
augmenting the input embedding with a topology-aware encoder (e.g., graph neural networks) to fuse explicit 
structural representations with sequence tokens is another promising extension. Identifying the optimal model 
scale for characterizing power system dynamics remains an open question and warrants further investigation.}


\appendix

{\color{red}
\section{Generic Feature Extraction of T-Blocks}\label{Generic Feature Extraction of Transformer Block}
{\color{red}
The decision to freeze the T-blocks in GPT is grounded in their optimization during pre-training, which shapes them into generic sequence feature extractors. The self-attention input and mapping are first defined, followed by the gradient upper-bound result.

For a given channel index $i$ and sample index $j$, we follow the notation in the main text and denote the input to the T-block as the embedded token matrix $\mathbf{z}_{ij}^{0}$ defined in \eqref{input embedding}. To simplify the analysis below, we rewrite it as
\begin{align}
\mathbf{X}=\mathbf{z}_{ij}^{0}=(x_1,\ldots,x_P)^\top \in \mathbb{R}^{P\times d},
\end{align}
where $P$ is the number of temporal patch tokens and, by a slight abuse of notation for notational simplicity, $x_m\in\mathbb{R}^{d}$ denotes the embedding of the $m$-th patch token (i.e., the $m$-th row of $\mathbf{z}_{ij}^{0}$). Define the self-attention mapping as
\begin{align}
f(\mathbf{X})=\mathrm{softmax}\!\left(\mathbf{X}A\mathbf{X}^\top\right)\mathbf{X},
\end{align}
where $A=\mathbf{W}^{Q}(\mathbf{W}^{K})^\top\in\mathbb{R}^{d\times d}$. Let $\mathbf{P}=\mathrm{softmax}(\mathbf{X}A\mathbf{X}^\top)$ with entries
\begin{align}
p_{m,n}=\frac{\exp(x_m^\top A x_n)}{\sum_{k=1}^{P}\exp(x_m^\top A x_k)}.
\end{align}

Based on the gradient analysis of self-attention in \cite{zhou2023one, kim2021lipschitz}, the following bound holds.

\noindent\textbf{Lemma A.1.}
Let $G=\left[\frac{\partial f(\mathbf{X})}{\partial x_{m}}\right]_{m=1}^{P}$ denote the Jacobian of $f(\mathbf{X})$ with respect to $\mathbf{X}$. Then,
\begin{align}
\begin{aligned}
|G|_2 \leq\;& |A|_2\sum_{m}\left(p_{m,m}+\frac{1}{2}\right)\underbrace{\left|x_m-\sum_{n}p_{m,n}x_n\right|^2}_{\text{Self-alignment}}\\
&+|A|_2\sum_{m\neq n}p_{m,n}\underbrace{\left|x_n-\sum_{k}p_{m,k}x_k\right|^2}_{\text{Cross-alignment}}
+\frac{|A|_2}{2}\sum_m|x_m|^2.
\label{upper bound of attention mechanism}
\end{aligned}
\end{align}

Equation \eqref{upper bound of attention mechanism} provides a direct objective-level explanation for the effect of pre-training. During large-scale pre-training, parameter updates in self-attention tend to reduce the gradient magnitude $|G|_2$; thus, optimization implicitly pushes down the right-hand side of \eqref{upper bound of attention mechanism}. This induces two structured alignment effects.

\begin{itemize}
\item First, the \textbf{Self-alignment} term
$\left|x_m-\sum_{n}p_{m,n}x_n\right|^2$
shrinks the distance between each token representation $x_m$ and its attention-weighted context
$\bar{x}_m=\sum_{n}p_{m,n}x_n$,
which improves feature consistency after repeated T-block transformations.
\item Second, the \textbf{Cross-alignment} term
$\left|x_n-\sum_{k}p_{m,k}x_k\right|^2$
is weighted by $p_{m,n}$. Therefore, token pairs with larger attention weights are more strongly constrained to align with the same contextual center $\sum_{k}p_{m,k}x_k$, yielding co-directional representations for strongly correlated tokens.
\end{itemize}

Together, these alignment dynamics tend to project token representations onto a compact and consistent principal subspace. Since this mechanism operates on ordered token sequences without restricting the form of $X$, it is independent of semantic modality and can be regarded as modality-agnostic. For power system transient prediction, such structured representation is particularly beneficial, as transient dynamics are typically dominated by a limited number of principal oscillatory modes. The alignment-induced projection concentrates trajectory information onto these dominant directions while suppressing redundant components, thereby stabilizing long-range dependency modeling and mitigating error accumulation during iterative forecasting. This optimization-induced generic feature extraction behavior explains why freezing the pre-trained T-block parameters remains effective for TSA tasks.

}

}

\bibliographystyle{elsarticle-num}  
\bibliography{ref}

@article{transient_stability_definition,
  title={Definition and classification of power system stability--revisited \& extended},
  author={Hatziargyriou, Nikos and Milanovic, Jovica and Rahmann, Claudia and Ajjarapu, Venkataramana and Canizares, Claudio and Erlich, Istvan and Hill, David and Hiskens, Ian and Kamwa, Innocent and Pal, Bikash and others},
  journal={IEEE Transactions on Power Systems},
  volume={36},
  number={4},
  pages={3271--3281},
  year={2020},
  publisher={IEEE}
}

@article{zhao2022structure,
  title={Structure-informed graph learning of networked dependencies for online prediction of power system transient dynamics},
  author={Zhao, Tianqiao and Yue, Meng and Wang, Jianhui},
  journal={IEEE Transactions on Power Systems},
  volume={37},
  number={6},
  pages={4885--4895},
  year={2022},
  publisher={IEEE}
}

@article{cai2020data,
  title={A data-based learning and control method for long-term voltage stability},
  author={Cai, Huaxiang and Ma, Haomin and Hill, David J},
  journal={IEEE Transactions on Power Systems},
  volume={35},
  number={4},
  pages={3203--3212},
  year={2020},
  publisher={IEEE}
}

@article{ye2024use,
  title={The Use of Machine Learning for Prediction of Post-Fault Rotor Angle Trajectories},
  author={Ye, Xinlin and Radovanovic, Ana and Milanovic, Jovica V},
  journal={IEEE Transactions on Power Systems},
  volume={39},
  number={5},
  pages={6496--6507},
  year={2024},
  publisher={IEEE}
}

@inproceedings{liu2024timer,
  title={Timer: generative pre-trained transformers are large time series models},
  author={Liu, Yong and Zhang, Haoran and Li, Chenyu and Huang, Xiangdong and Wang, Jianmin and Long, Mingsheng},
  booktitle={Proceedings of the 41st International Conference on Machine Learning},
  pages={32369--32399},
  year={2024}
}

@article{radford2019language,
  title={Language models are unsupervised multitask learners},
  author={Radford, Alec and Wu, Jeffrey and Child, Rewon and Luan, David and Amodei, Dario and Sutskever, Ilya and others},
  journal={OpenAI blog},
  volume={1},
  number={8},
  pages={9},
  year={2019}
}

@article{vaswani2017attention,
  title={Attention is all you need},
  author={Vaswani, Ashish and Shazeer, Noam and Parmar, Niki and Uszkoreit, Jakob and Jones, Llion and Gomez, Aidan N and Kaiser, {\L}ukasz and Polosukhin, Illia},
  journal={Advances in neural information processing systems},
  volume={30},
  year={2017}
}

@article{zhou2023one,
  title={One fits all: Power general time series analysis by pretrained lm},
  author={Zhou, Tian and Niu, Peisong and Sun, Liang and Jin, Rong and others},
  journal={Advances in neural information processing systems},
  volume={36},
  pages={43322--43355},
  year={2023}
}

@article{qiu2022adaptive,
  title={Adaptive Lyapunov function method for power system transient stability analysis},
  author={Qiu, Zitian and Duan, Chao and Yao, Wei and Zeng, Pingliang and Jiang, Lin},
  journal={IEEE Transactions on Power Systems},
  volume={38},
  number={4},
  pages={3331--3344},
  year={2022},
  publisher={IEEE}
}

@article{shen2025physics,
  title={Physics-following Neural Network for Online Dynamic Security Assessment},
  author={Shen, Chao and Zuo, Ke and Sun, Mingyang},
  journal={IEEE Transactions on Power Systems},
  year={2025},
  publisher={IEEE}
}

@article{ren2021interpretable,
  title={An interpretable deep learning method for power system transient stability assessment via tree regularization},
  author={Ren, Chao and Xu, Yan and Zhang, Rui},
  journal={IEEE Transactions on Power Systems},
  volume={37},
  number={5},
  pages={3359--3369},
  year={2021},
  publisher={IEEE}
}

@article{chen2022interpretable,
  title={Interpretable time-adaptive transient stability assessment based on dual-stage attention mechanism},
  author={Chen, Qifan and Lin, Nan and Bu, Siqi and Wang, Huaiyuan and Zhang, Baohui},
  journal={IEEE Transactions on Power Systems},
  volume={38},
  number={3},
  pages={2776--2790},
  year={2022},
  publisher={IEEE}
}

@article{gomez2010support,
  title={Support vector machine-based algorithm for post-fault transient stability status prediction using synchronized measurements},
  author={Gomez, Francisco R and Rajapakse, Athula D and Annakkage, Udaya D and Fernando, Ioni T},
  journal={IEEE Transactions on Power systems},
  volume={26},
  number={3},
  pages={1474--1483},
  year={2010},
  publisher={IEEE}
}

@ARTICLE{Sun2007An,
  author={Sun, Kai and Likhate, Siddharth and Vittal, Vijay and Kolluri, V. Sharma and Mandal, Sujit},
  journal={IEEE Transactions on Power Systems}, 
  title={An Online Dynamic Security Assessment Scheme Using Phasor Measurements and Decision Trees}, 
  year={2007},
  volume={22},
  number={4},
  pages={1935-1943},
  doi={10.1109/TPWRS.2007.908476}}

@article{zhou1994application,
  title={Application of artificial neural networks in power system security and vulnerability assessment},
  author={Zhou, Qin and Davidson, Jennifer and Fouad, AA},
  journal={IEEE Transactions on Power Systems},
  volume={9},
  number={1},
  pages={525--532},
  year={1994},
  publisher={IEEE}
}

@article{azman2020unified,
  title={A unified online deep learning prediction model for small signal and transient stability},
  author={Azman, Syafiq Kamarul and Isbeih, Younes J and El Moursi, Mohamed Shawky and Elbassioni, Khaled},
  journal={IEEE transactions on power systems},
  volume={35},
  number={6},
  pages={4585--4598},
  year={2020},
  publisher={IEEE}
}

@article{zhu2023integrated,
  title={Integrated data-driven power system transient stability monitoring and enhancement},
  author={Zhu, Lipeng and Wen, Weijia and Li, Jiayong and Hu, Yuhan},
  journal={IEEE Transactions on Power Systems},
  volume={39},
  number={1},
  pages={1797--1809},
  year={2023},
  publisher={IEEE}
}

@article{su2023online,
  title={Online transient stability margin estimation using improved deep learning ensemble model},
  author={Su, Heng-Yi and Lai, Chia-Ching},
  journal={IEEE Transactions on Power Systems},
  volume={39},
  number={6},
  pages={7421--7424},
  year={2023},
  publisher={IEEE}
}

@article{zhu2019hierarchical,
  title={Hierarchical deep learning machine for power system online transient stability prediction},
  author={Zhu, Lipeng and Hill, David J and Lu, Chao},
  journal={IEEE Transactions on Power Systems},
  volume={35},
  number={3},
  pages={2399--2411},
  year={2019},
  publisher={IEEE}
}

@article{zhu2021networked,
  title={Networked time series shapelet learning for power system transient stability assessment},
  author={Zhu, Lipeng and Hill, David J},
  journal={IEEE Transactions on Power Systems},
  volume={37},
  number={1},
  pages={416--428},
  year={2021},
  publisher={IEEE}
}

@inproceedings{misyris2020physics,
  title={Physics-informed neural networks for power systems},
  author={Misyris, George S and Venzke, Andreas and Chatzivasileiadis, Spyros},
  booktitle={2020 IEEE power \& energy society general meeting (PESGM)},
  pages={1--5},
  year={2020},
  organization={IEEE}
}

@article{cui2023frequency,
  title={A frequency domain approach to predict power system transients},
  author={Cui, Wenqi and Yang, Weiwei and Zhang, Baosen},
  journal={IEEE Transactions on Power Systems},
  volume={39},
  number={1},
  pages={465--477},
  year={2023},
  publisher={IEEE}
}

@article{tan2025bayesian,
  title={Bayesian Post-Fault Power System Dynamic Trajectory Prediction},
  author={Tan, Bendong and Zhao, Junbo},
  journal={IEEE Transactions on Power Systems},
  year={2025},
  publisher={IEEE}
}

@article{tu2024powerpm,
  title={Powerpm: Foundation model for power systems},
  author={Tu, Shihao and Zhang, Yupeng and Zhang, Jing and Fu, Zhendong and Zhang, Yin and Yang, Yang},
  journal={Advances in Neural Information Processing Systems},
  volume={37},
  pages={115233--115260},
  year={2024}
}

@inproceedings{kim2021lipschitz,
  title={The lipschitz constant of self-attention},
  author={Kim, Hyunjik and Papamakarios, George and Mnih, Andriy},
  booktitle={International Conference on Machine Learning},
  pages={5562--5571},
  year={2021},
  organization={PMLR}
}

@article{lu2024advanced,
  title={Advanced Probabilistic Transient Stability Assessment for Operational Planning: A Physics-Informed Graphical Learning Approach},
  author={Lu, Genghong and Bu, Siqi},
  journal={IEEE Transactions on Power Systems},
  year={2024},
  publisher={IEEE}
}

@article{lamb2016professor,
  title={Professor forcing: A new algorithm for training recurrent networks},
  author={Lamb, Alex M and ALIAS PARTH GOYAL, Anirudh Goyal and Zhang, Ying and Zhang, Saizheng and Courville, Aaron C and Bengio, Yoshua},
  journal={Advances in neural information processing systems},
  volume={29},
  year={2016}
}

@article{achiam2023gpt,
  title={Gpt-4 technical report},
  author={Achiam, Josh and Adler, Steven and Agarwal, Sandhini and Ahmad, Lama and Akkaya, Ilge and Aleman, Florencia Leoni and Almeida, Diogo and Altenschmidt, Janko and Altman, Sam and Anadkat, Shyamal and others},
  journal={arXiv preprint arXiv:2303.08774},
  year={2023}
}

@article{liu2024deepseek,
  title={Deepseek-v3 technical report},
  author={Liu, Aixin and Feng, Bei and Xue, Bing and Wang, Bingxuan and Wu, Bochao and Lu, Chengda and Zhao, Chenggang and Deng, Chengqi and Zhang, Chenyu and Ruan, Chong and others},
  journal={arXiv preprint arXiv:2412.19437},
  year={2024}
}

@article{shen2026proopf,
  title={ProOPF: Benchmarking and Improving LLMs for Professional-Grade Power Systems Optimization Modeling},
  author={Shen, Chao and Guo, Zihan and Wan, Xu and Yang, Zhenghao and Zhang, Yifan and Huang, Wengi and Song, Jie and Zhang, Zongyan and Sun, Mingyang},
  journal={arXiv preprint arXiv:2602.03070},
  year={2026}
}

@article{shen2025physics_argumented,
  title={Physics-Augmented Auxiliary Learning for Power System Transient Stability Assessment},
  author={Shen, Chao and Zuo, Ke and Sun, Mingyang},
  journal={IEEE Transactions on Industrial Informatics},
  year={2025},
  publisher={IEEE}
}

@article{segurado2011increasing,
  title={Increasing the penetration of renewable energy resources in S. Vicente, Cape Verde},
  author={Segurado, Raquel and Kraja{\v{c}}i{\'c}, Goran and Dui{\'c}, Neven and Alves, Lu{\'\i}s},
  journal={Applied energy},
  volume={88},
  number={2},
  pages={466--472},
  year={2011},
  publisher={Elsevier}
}

@article{shi2020convolutional,
  title={Convolutional neural network-based power system transient stability assessment and instability mode prediction},
  author={Shi, Zhongtuo and Yao, Wei and Zeng, Lingkang and Wen, Jianfeng and Fang, Jiakun and Ai, Xiaomeng and Wen, Jinyu},
  journal={Applied Energy},
  volume={263},
  pages={114586},
  year={2020},
  publisher={Elsevier}
}

@article{zhao2019exergy,
  title={Exergy analysis of the regulating measures of operational flexibility in supercritical coal-fired power plants during transient processes},
  author={Zhao, Yongliang and Liu, Ming and Wang, Chaoyang and Wang, Zhu and Chong, Daotong and Yan, Junjie},
  journal={Applied Energy},
  volume={253},
  pages={113487},
  year={2019},
  publisher={Elsevier}
}

@article{zhan2023hybrid,
  title={A hybrid transfer learning method for transient stability prediction considering sample imbalance},
  author={Zhan, Xianwen and Han, Song and Rong, Na and Cao, Yun},
  journal={Applied Energy},
  volume={333},
  pages={120573},
  year={2023},
  publisher={Elsevier}
}

@article{zou2020least,
  title={A least-squares support vector machine method for modeling transient voltage in polymer electrolyte fuel cells},
  author={Zou, Wei and Froning, Dieter and Shi, Yan and Lehnert, Werner},
  journal={Applied energy},
  volume={271},
  pages={115092},
  year={2020},
  publisher={Elsevier}
}

@article{li2022deep,
  title={A deep-learning intelligent system incorporating data augmentation for short-term voltage stability assessment of power systems},
  author={Li, Yang and Zhang, Meng and Chen, Chen},
  journal={Applied Energy},
  volume={308},
  pages={118347},
  year={2022},
  publisher={Elsevier}
}

@article{fan2019assessment,
  title={Assessment of deep recurrent neural network-based strategies for short-term building energy predictions},
  author={Fan, Cheng and Wang, Jiayuan and Gang, Wenjie and Li, Shenghan},
  journal={Applied energy},
  volume={236},
  pages={700--710},
  year={2019},
  publisher={Elsevier}
}

@article{karimi2022optimization,
  title={Optimization-driven uncertainty forecasting: Application to day-ahead commitment with renewable energy resources},
  author={Karimi, Sajad and Kwon, Soongeol},
  journal={Applied Energy},
  volume={326},
  pages={119929},
  year={2022},
  publisher={Elsevier}
}

@article{gao2025mmgpt4lf,
  title={MMGPT4LF: Leveraging an optimized pre-trained GPT-2 model with multi-modal cross-attention for load forecasting},
  author={Gao, Mingyang and Zhou, Suyang and Gu, Wei and Wu, Zhi and Liu, Haiquan and Zhou, Aihua and Wang, Xinliang},
  journal={Applied Energy},
  volume={392},
  pages={125965},
  year={2025},
  publisher={Elsevier}
}

@article{wen2016allocation,
  title={Allocation of ESS by interval optimization method considering impact of ship swinging on hybrid PV/diesel ship power system},
  author={Wen, Shuli and Lan, Hai and Hong, Ying-Yi and Yu, David C and Zhang, Lijun and Cheng, Peng},
  journal={Applied Energy},
  volume={175},
  pages={158--167},
  year={2016},
  publisher={Elsevier}
}

@article{shu2015investigation,
  title={Investigation of offshore wind energy potential in Hong Kong based on Weibull distribution function},
  author={Shu, ZR and Li, QS and Chan, PW},
  journal={Applied Energy},
  volume={156},
  pages={362--373},
  year={2015},
  publisher={Elsevier}
}

@article{xue1989extended,
  title={Extended equal area criterion justifications, generalizations, applications},
  author={Xue, Yusheng and Van Custem, Thierry and Ribbens-Pavella, Mania},
  journal={IEEE Transactions on Power Systems},
  volume={4},
  number={1},
  pages={44--52},
  year={1989},
  publisher={IEEE}
}

@article{kundur2007power,
  title={Power system stability},
  author={Kundur, Prabha and others},
  journal={Power system stability and control},
  volume={10},
  number={1},
  pages={7--1},
  year={2007}
}

@article{chang1995direct,
  title={Direct stability analysis of electric power systems using energy functions: theory, applications, and perspective},
  author={Chang, Hsiao-Dong and Chu, Chia-Chi and Cauley, Gerry},
  journal={Proceedings of the IEEE},
  volume={83},
  number={11},
  pages={1497--1529},
  year={1995},
  publisher={IEEE}
}

@article{berggren2002nature,
  title={On the nature of unstable equilibrium points in power systems},
  author={Berggren, B and Andersson, G},
  journal={IEEE transactions on power systems},
  volume={8},
  number={2},
  pages={738--745},
  year={2002},
  publisher={IEEE}
}

@article{zong2023transient,
  title={Transient stability assessment of large-scale power system using predictive maximal Lyapunov exponent approach},
  author={Zong, Qihang and Yao, Wei and Zhou, Hongyu and Zhao, Haiyu and Wen, Jinyu and Cheng, Shijie},
  journal={IEEE Transactions on Power Systems},
  volume={39},
  number={3},
  pages={5163--5176},
  year={2023},
  publisher={IEEE}
}

@article{willems2007application,
  title={The application of Lyapunov methods to the computation of transient stability regions for multimachine power systems},
  author={Willems, Jacques L and Willems, Jan C},
  journal={IEEE Transactions on Power Apparatus and Systems},
  number={5},
  pages={795--801},
  year={2007},
  publisher={IEEE}
}

@article{chiang1988pebs,
  title={Foundations of the potential energy boundary surface method for power system transient stability analysis},
  author={Chiang, H.-D. and Wu, F. F. and Varaiya, P. P.},
  journal={IEEE Transactions on Circuits and Systems},
  volume={35},
  number={6},
  pages={712--728},
  year={1988},
  doi={10.1109/31.1808},
  publisher={IEEE}
}

@misc{yu2025dapoopensourcellmreinforcement,
  title={DAPO: An Open-Source LLM Reinforcement Learning System at Scale},
  author={Yu, Qiying and Zhang, Zheng and Zhu, Ruofei and Yuan, Yufeng and Zuo, Xiaochen and Yue, Yu and Dai, Weinan and Fan, Tiantian and Liu, Gaohong and Liu, Lingjun and Liu, Xin and Lin, Haibin and Lin, Zhiqi and Ma, Bole and Sheng, Guangming and Tong, Yuxuan and Zhang, Chi and Zhang, Mofan and Zhang, Wang and Zhu, Hang and Zhu, Jinhua and Chen, Jiaze and Chen, Jiangjie and Wang, Chengyi and Yu, Hongli and Song, Yuxuan and Wei, Xiangpeng and Zhou, Hao and Liu, Jingjing and Ma, Wei-Ying and Zhang, Ya-Qin and Yan, Lin and Qiao, Mu and Wu, Yonghui and Wang, Mingxuan},
  year={2025},
  eprint={2503.14476},
  archivePrefix={arXiv},
  primaryClass={cs.LG},
  url={https://arxiv.org/abs/2503.14476}
}

@article{shen2026llm,
  title={LLM-DMD: Large Language Model-based Power System Dynamic Model Discovery},
  author={Shen, Chao and Guo, Zihan and Zuo, Ke and Huang, Wenqi and Sun, Mingyang},
  journal={arXiv preprint arXiv:2601.05632},
  year={2026}
}

\end{document}